\documentclass[aps,floats, floatfix, superscriptaddress,tightenlines,titlepage, nofootinbib]{revtex4-1}
\usepackage{amsmath, amssymb, amsfonts, amsthm, latexsym, epsfig, mathrsfs, xcolor, bbm, slashed}
\usepackage[top=1in, bottom=1in, left=.9in, right=.9in]{geometry}
\usepackage{mdframed}
\DeclareMathAlphabet{\mathpzc}{OT1}{pzc}{m}{it}
\usepackage[T1]{fontenc} 
\usepackage{amsfonts}
\usepackage{graphicx}
\usepackage{amsmath}
\usepackage{color}
\usepackage{mathtools}

\usepackage[colorlinks=true, citecolor=blue,urlcolor=blue]{hyperref}
\usepackage{hyperref}
\usepackage{subfigure}
\usepackage{comment}
\usepackage{multirow}
\usepackage{calligra}

\newcommand{\AR}[1]{\textcolor{cyan}{\textsf{[AR: #1]}}}
\newcommand{\SG}[1]{\textcolor{blue}{\textsf{[SG: #1]}}}

\newcommand{\xh}{Y}

\def\pd{\partial }

\begin{document}

\title{Horizon Instability of the Extremal BTZ Black Hole}

\author{Samuel E.~Gralla}
\email{sgralla@email.arizona.edu}
\affiliation{Department of Physics, University of Arizona}

\author{Arun~Ravishankar}
\email{arunravishankar@email.arizona.edu}
\affiliation{Department of Physics, University of Arizona}

\author{Peter~Zimmerman}
\email{peterzimmerman@aei.mpg.de}
\affiliation{Max Planck Institute for Gravitational Physics (Albert Einstein Institute)\\
  Am M\"uhlenberg 1, 14476 Potsdam, Germany}

\date{\today}
\begin{abstract}
We study real-time propagation of a massive scalar field on the extremal BTZ black hole spacetime, focusing on the Aretakis instability of the event horizon.  We obtain a simple time-domain expression for the $\textrm{AdS}_3$ retarded Green function with Dirichlet boundary conditions and construct the corresponding time-domain BTZ retarded Green function using the method of images.  The field decays at different rates on and off the horizon, indicating that transverse derivatives grow with time on the horizon (Aretakis instability).  We solve the null geodesic equation in full generality and show that the instability is associated with a class of null geodesics that orbit near the event horizon arbitrarily many times before falling in.  In an appendix we also treat the problem in the frequency domain, finding consistency between the methods. 

\end{abstract}
\maketitle
{
  \hypersetup{linkcolor=black}
  \tableofcontents
}
\newpage
\section{Introduction}
In the decade since Aretakis' initial study of massless scalar fields in the extremal Reissner-N{\"o}rdstrom spacetime \cite{Aretakis:2010gd,Aretakis:2011hc}, it has become clear that extremal horizons generically exhibit weak derivative instabilities \cite{Aretakis:2012ei,Lucietti:2012xr,Gralla:2018xzo}.  Independent of the type of extremal black hole (and even of the type of field that perturbs it!), the evidence suggests that sufficiently high-order transverse derivatives \textit{always} grow at least polynomially in advanced time along the event horizon.  This has the physical consequence that infalling observers experience large field  gradients \cite{Lucietti:2012xr,Gralla:2017lto,Gralla:2016sxp}.

This general picture has emerged from a variety of techniques, whose domains of validity are largely non-overlapping.  The original technique of Aretakis \cite{Aretakis:2012ei,Aretakis:2012bm,Aretakis:2013oja} involves a conserved quantity along the horizon that provides an obstruction to decay.  This technique has been used only in what we call the ``discrete case'' (e.g., axisymmetric massless perturbations of Kerr \cite{Gralla:2018xzo}), and mainly when initial data extends to the horizon.  We ourselves have been involved with a different technique based on frequency-domain analysis, in which the instability is associated with a singular branch point in each mode of the retarded Green function \cite{Leaver1986,Casals:2016mel,Gralla:2017lto,Casals:2018eev}. This technique has been used mainly in the \textit{non-}discrete case (e.g., non-axisymmetric mode perturbations of Kerr \cite{Gralla:2017lto}), and only when initial data is confined \textit{away} from the horizon.  Other techniques rely on discrete inversion symmetries present only for some black holes \cite{Bizon:2012we,Lucietti:2012xr,Godazgar:2017igz, Bhattacharjee:2018pqb}. Analytic arguments involving the near-horizon geometry \cite{Lucietti:2012sf,Hadar:2017ven,Gralla:2018xzo,Gralla:2017lto} together with numerical work in a variety of settings \cite{Lucietti:2012xr,Murata:2013daa,Burko:2017eky}, have provided complimentary insight into these various regimes.  However, a fully general understanding of the instability remains elusive.

There is a dissonance here: a universal phenomenon should be simple at its core, yet the state of the art presents a sprawling complexity.  One suspects that a key unifying idea has not yet been identified.  In such a situation, it pays to study the effect in the simplest possible setting, where the nature of the phenomenon might be revealed with a minimum of distraction.  Buoyed by this hope, in this paper we initiate the study of the Aretakis instability in the exceptionally simple setting of a 2+1-dimensional black hole---the Ba\~{n}ados-Teitelboim-Zanelli (BTZ) spacetime \cite{Banados:1992gq,Banados:1992wn}.  As this black hole has negative cosmological constant, the results may also help understand the holographic implications of the instability \cite{Gralla:2018xoz,Hadar:2018izi}.

The BTZ black hole allows exceptional analytic control because it is locally isometric to three-dimensional Anti-de Sitter space ($\textrm{AdS}_3$), a maximally symmetric spacetime.  We focus on the retarded Green function of a massive scalar field satisfying Dirichlet boundary conditions.  Using the results of \cite{Danielsson:1998wt} (and correcting some minor computational errors), we write the $\textrm{AdS}_3$ retarded Green function in a very simple form.  We then use the method of images \cite{Steif:1993zv} to construct the BTZ Green function.

The Aretakis instability emerges from this image sum in a beautiful way.  Each term in the sum decays at the same rate, fixed by the $\textrm{AdS}_3$ spacetime.  For a field point off the horizon in BTZ, the high-order images are unimportant, and the full Green function (and hence field) also decays at this $\textrm{AdS}_3$ rate.  However, when the field point is chosen on the BTZ horizon, the high-order images become important and the infinite sum decays at precisely \textit{half} the $\textrm{AdS}_3$ rate,\footnote{We demonstrate this fact numerically and give heuristic analytical arguments for the behavior, but do not provide a rigorous proof.  Furthermore, this analysis assumes that the source point is outside the horizon, or equivalently that the initial data does not extend to the horizon.} as expected on general grounds \cite{Gralla:2018xzo}.  Since the field decays at different rates on and off the horizon, sufficiently high-order transverse derivatives on the horizon must grow---the Aretakis instability.  

By solving the null geodesic equation in BTZ, we are able to identify the late-time-relevant terms in the image sum with the arrival of wavefronts that have orbited the event horizon arbitrarily many times before falling in.  The number of orbits increases linearly as the geodesic conserved quantity approaches that of the event horizon.  We conjecture that this behavior is part of the following larger pattern for black holes: The surface gravity of the event horizon functions as a Lyapunov exponent for the deviation of nearby null geodesics; and when this exponent vanishes for an extremal black hole, the exponential deviation goes over to a power law.  One can then say that wavefronts linger far longer near the event horizon of extremal black holes, as compared to analogous non-extremal black holes, providing an interpretation of the instability of extremal horizons.


Our results offer a new way to think about the Aretakis instability, but they do not solve the problem of unifying the previous different approaches.  To provide a point of comparison, in the appendices we apply two other methods to the extremal BTZ spacetime: the conserved charge method (App.~\ref{sec:Aretakis}) and the mode sum method (App.~\ref{sec:modes}).  We cannot compare with the conserved charged method because our analysis is restricted to the non-discrete case, with initial data not extending to the horizon.  We can compare with the decay rate of individual modes in the frequency domain analysis (and we find agreement), but this method does not give the behavior of the full field (see further discussion in App.~\ref{sec:modes}).  Thus there remains much mystery surrounding the Aretakis instability of extremal horizons.

This paper is organized as follows. In Sec.~\ref{sec:metric} we review the $\textrm{AdS}_3$ and extremal BTZ spacetimes, introducing notation.  In Sec.~\ref{sec:scalar} we consider a massive scalar field and construct the retarded Green function by the method of images.  In Sec.~\ref{sec:late} we discuss late-time decay, and in Sec.~\ref{sec:geodesics} we interpret the instability in terms of null geodesics.  Our signature is $(-++)$ and we set the $\text{AdS}$ radius to one.

\section{Metric}\label{sec:metric}

We now discuss the $\textrm{AdS}_3$ metric and its periodic identification to produce the extremal BTZ black hole.  We introduce a few different coordinate patches that are useful in the analysis that follows.

\subsection{Global $\textrm{AdS}_3$}
The maximally extended $\text{AdS}_{3}$ metric is given in ``global coordinates'' by
\begin{equation}
    ds^2 = \frac{1}{\cos^2{\chi}} \left(-d\tau^2 + d\chi^2 + \sin^{2}{\chi} d \Omega^2\right),
\end{equation}
where $\tau\in(-\infty,\infty)$, $\chi \in [0, \pi/2)$, and $\Omega \sim \Omega+2\pi$.  The spacetime can be conformally completed by a cylinder at $\chi = \pi/2$, which we refer to as the boundary.  We will generally use global coordinates to visualize the spacetime, with $(\tau,\chi,\Omega)$ treated as cylindrical coordinates representing height, radius, and angle, respectively.

\begin{figure}
\centering
  \includegraphics[width=.32\linewidth]{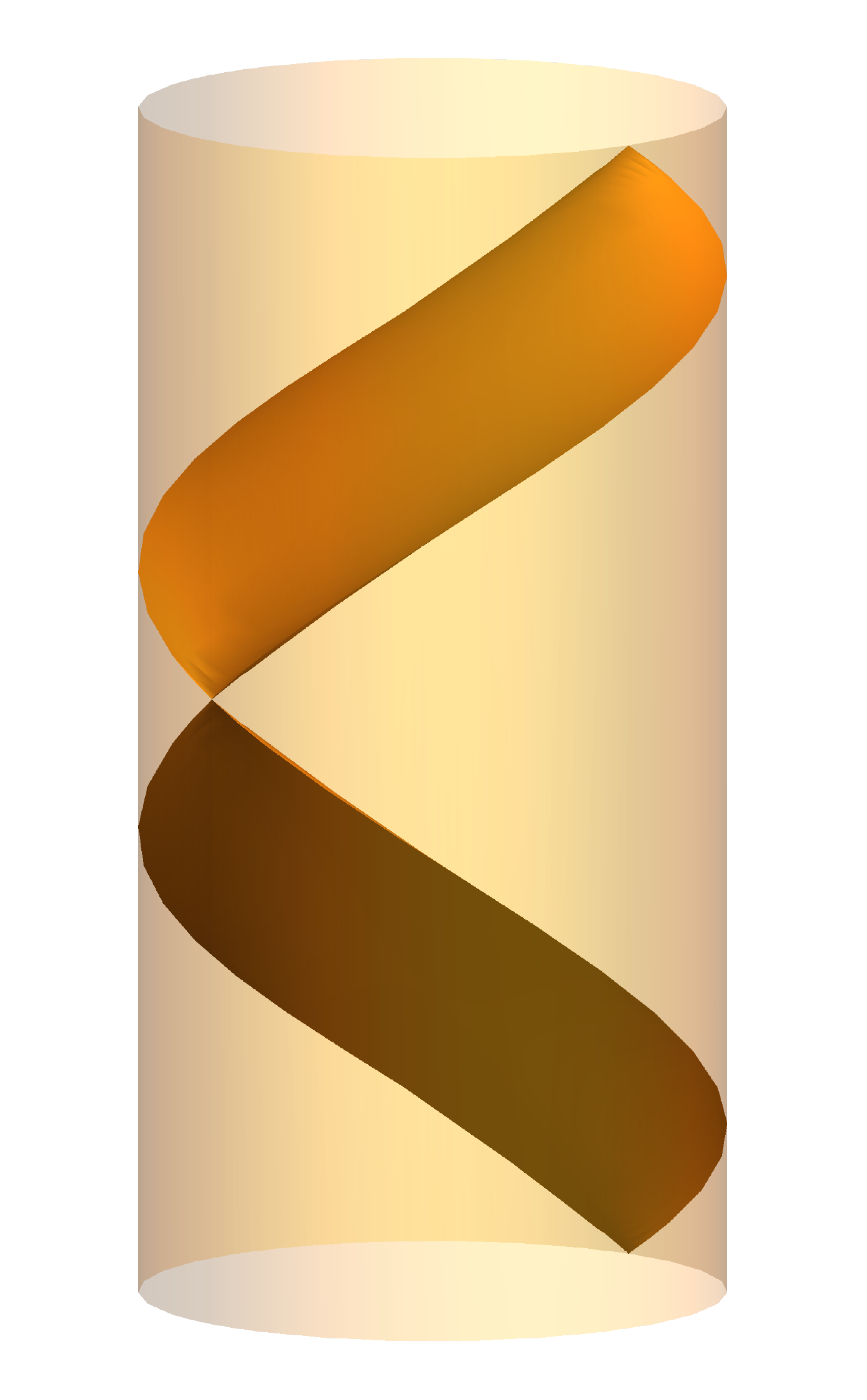}
  \includegraphics[width=.32\linewidth]{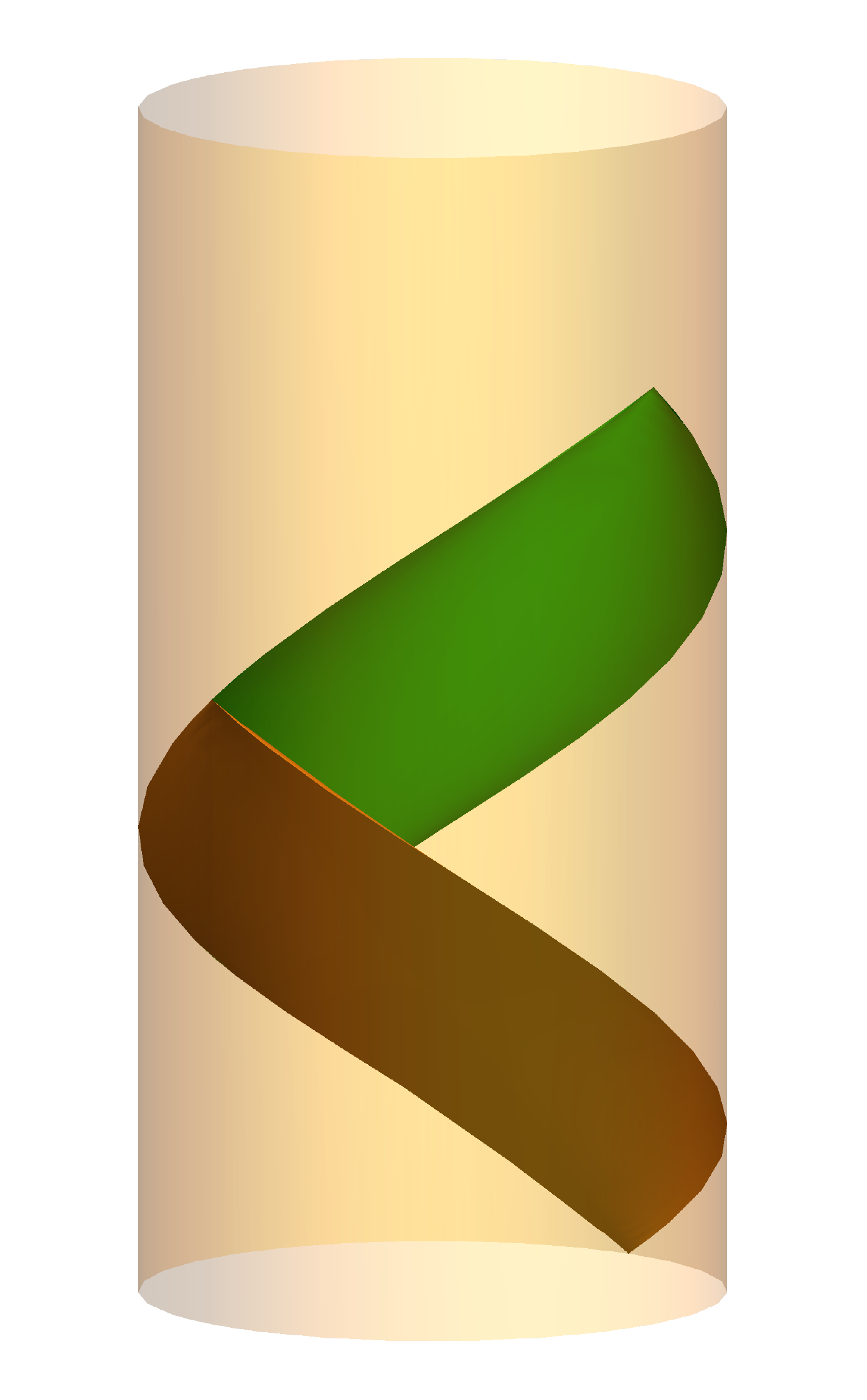}
    \includegraphics[width=.32\linewidth]{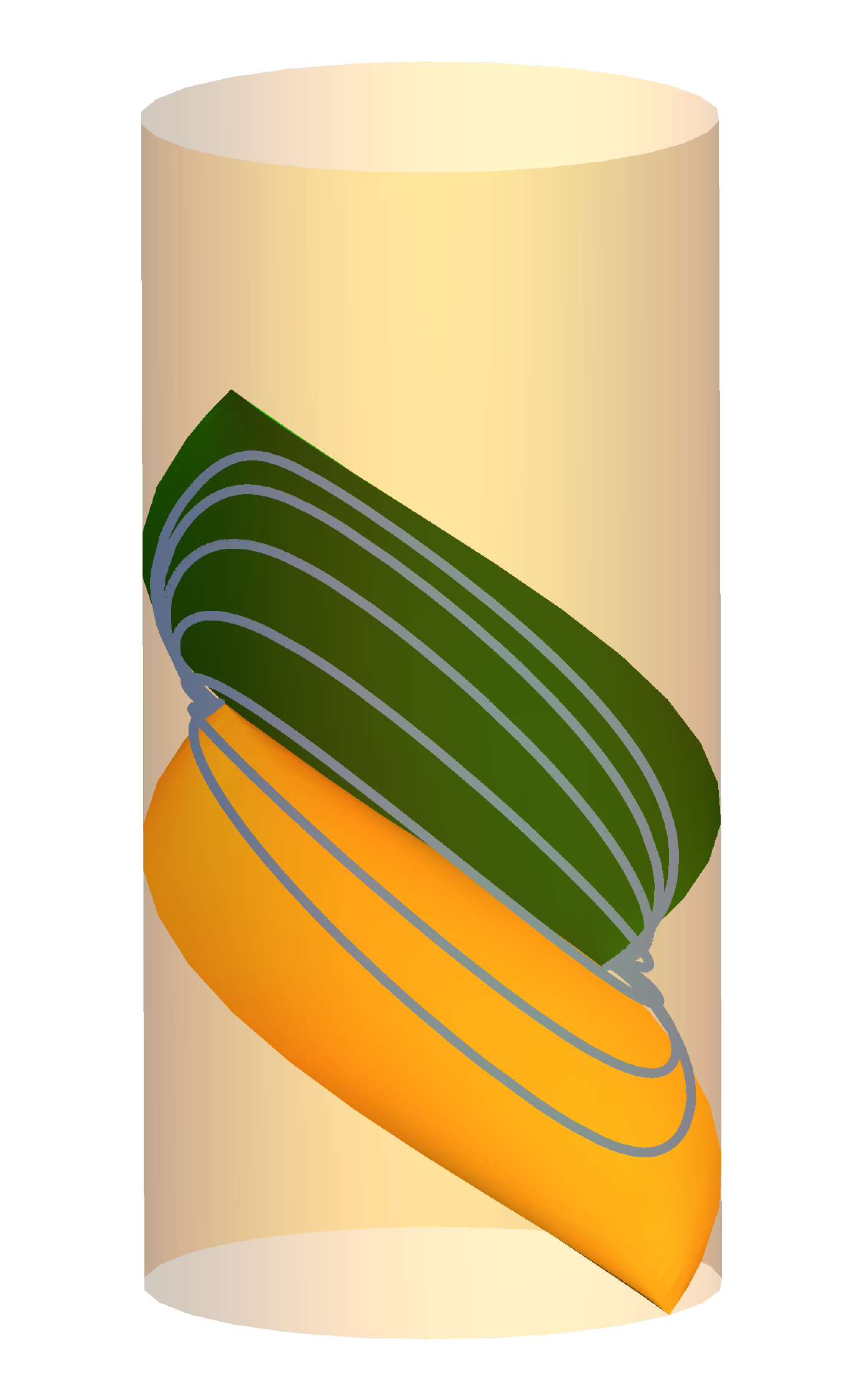}
  \caption{$\text{AdS}_3$ and its patches, plotted using global coordinates $(\tau,\chi,\Omega)$ as cylindrical coordinates with height $\tau$, radius $\chi$, and angle $\Omega$. The maximally extended spacetime has a cylindrical conformal boundary at $\chi=\pi/2$, shown here in translucent orange.  On the left, we show the Poincar\'e horizons $\tau_{\pm}$ that bound the Poincar\'e patch.  In the middle, we show $\tau_-$ along with the future horizon $\tau_H$ (which is also a Poincar\'e horizon); these bound the ``extremal patch'' that becomes the BTZ exterior after identification.  Finally, on the right we show the same plot from a different perspective, along with a selection of integral curves of the Killing field $\xi = \partial_{X}$.  The extremal BTZ black hole is obtained by identifying discretely along these curves.  
  }\label{fig:patches}
  \end{figure}

\subsection{Poincar\'e coordinates and patch}

 ``Poincar\'e coordinates'' for $\textrm{AdS}_3$ are given by
\begin{equation}
    t = \frac{\sin{\tau}}{\cos{\tau} - \cos{\Omega} \sin{\chi}}, \qquad z = \frac{\cos{\chi}}{\cos{\tau} - \cos{\Omega} \sin{\chi}}, \qquad    x = \frac{\sin{\Omega} \sin{\chi}}{\cos{\tau} - \cos{\Omega}\sin{\chi}}, \end{equation}
where the metric becomes
\begin{equation}
    ds^2 = \frac{1}{z^2} (-dt^2 + dz^2 + d x^2).
\end{equation}
These coordinates cover only a 
 ``Poincar\'e patch'' where $(\cos \tau - \cos \Omega \sin \chi)$ has a definite sign.  We choose the region $\tau \in (\tau_-,\tau_+)$, with
\begin{equation} \label{poincare-patch}
  \tau_{\pm}  := \pm \arccos{(\cos{\Omega}\sin{\chi})}.
\end{equation}
The bounding null surfaces $\tau=\tau_{\pm}$ are called Poincar\'e horizons (Fig.~\ref{fig:patches} left).  The interior has $z>0$ with $t$ and $x$ unbounded.  The boundary is at $z=0$, and all coordinates become large as the Poincar\'e horizons are approached.

\subsection{Extremal coordinates and patch}


We also introduce ``extremal coordinates'' for $\text{AdS}_3$, 
\begin{subequations}
\begin{align}\label{poinctoextpat}
t&=\frac{1}{2} \left( T+X-\frac{R_{H}}{R^2 - R_{H}^2} - \frac{e^{2R_{H}(X-T)}}{2R_{H}}\right) \\ 
z&= \frac{e^{R_{H}(X-T)}}{\sqrt{R^2 - R_{H}^2}} \\
x&= \frac{1}{2} \left( T+X-\frac{R_{H}}{R^2 - R_{H}^2} + \frac{e^{2R_{H}(X-T)}}{2R_{H}}\right),
\end{align}
\end{subequations}
bringing the metric into the form
\begin{equation}\label{extpat}
ds^2 = -(R^2 - 2R_{H}^2)dT^2 + \frac{R^2}{(R^2 - R_{H}^2)^2} dR^2  - 2R_{H}^2 dT dX + R^2 dX^2.
\end{equation}
These coordinates cover the region $R>R_H$, which is bounded by the past Poincar\'e horizon $\tau_-$ and a globally translated and rotated Poincar\'e horizon $\tau_H$ defined by\footnote{In global coordinates, $R>R_H$ corresponds to $\alpha \beta>0$ with $\alpha = \sin{\Omega} \sin{\chi} - \sin{\tau}$ and $\beta = \cos{\tau} - \cos{\Omega} \sin{\chi}$.  The zeros of $\beta$ bound a set of Poincar\'e patches including our choice \eqref{poincare-patch}, while the zeros of $\alpha$ bound an interleaving set related by the symmetries $\tau \to \tau+\pi/2$ and $\phi \to \phi + \pi/2$.  As our Poincar\'e patch \eqref{poincare-patch} has $\beta>0$, we require $\alpha>0$ as well, giving rise to $\tau \in (\tau_-,\tau_H)$.}
\begin{align}
    \tau_H  = \arcsin{(\sin{\Omega} \sin{\chi})}.
\end{align}
We refer to this patch $\tau \in(\tau_-,\tau_H)$ as the extremal patch (Fig.~\ref{fig:patches} middle and right). We call the bounding null surfaces $\tau_-$ and $\tau_H$ the past and future horizons (respectively), since these will become the event horizons of the extremal BTZ black hole after the identification $X \sim X+2\pi$.  These horizons are described by $R=R_H$ in extremal coordinates, but the patch itself is independent of $R_H$.\footnote{Note that the original transformation given by BTZ does entail a different patch for each value of $R_H$; these patches approach Poincar\'e patch as $R_H \to 0$.  Our extremal patch remains distinct in the $R_H \to 0$ limit.}  
The inverse transformation is given by 
\begin{subequations}
\begin{align}
    T &= \frac{2x^2 - 2t^2 + z^2}{4(x-t)} -\frac{1}{4R_{H}} \log{(2R_{H}(x-t))} \\
    R &= R_{H}\sqrt{1+ \frac{2(x-t)}{R_{H}z^2}}\\
    X &= \frac{2x^2 - 2t^2 + z^2}{4(x-t)} +\frac{1}{4R_{H}} \log{(2R_{H}(x-t))}.
\end{align}
\end{subequations}
\subsection{Horizon coordinates}

Finally, it will be useful to consider ``horizon coordinates'' ($V,R,\Phi$) defined by
\begin{subequations}\label{horizon-coordinates}
\begin{align}
    T &= V + \frac{R}{2 (R^2 -  R_{H}^2)}  - \frac{1}{4R_H}\,\log{\frac{R-R_{H}}{R + R_{H}}} \\
    X & =  \Phi  + V+ \frac{R}{2 (R^2 -  R_{H}^2)}  + \frac{1}{4R_H}\,\log{\frac{R-R_{H}}{R + R_{H}}},
\end{align}
\end{subequations}
where the metric becomes
\begin{align}\label{ingcormetric}
    ds^2 = 2dR dV + 2(R^2 - R_{H}^2)d \Phi d V + R^2 d \Phi^2.
\end{align}
These coordinates cover the future horizon $R=R_{H}$, with the null generators given by $\Phi=\rm{const}$.  When $\Phi \sim \Phi+2\pi$, these are ingoing, corotating coordinates for the extremal BTZ black hole.  The horizon-generating Killing field is given by
\begin{align}
    \zeta = \frac{\partial}{\partial V} = \frac{\partial}{\partial T} + \frac{\partial}{\partial X}.
\end{align}


\subsection{Extremal Killing field and BTZ}

The extremal BTZ black hole is obtained by identifying points separated by parameter distance $2 \pi$ along the integral curves of the Killing field 
\begin{align}
\xi = \frac{\pd}{\pd X} = \frac{\pd}{\pd \Phi}.
\end{align}
This means we identify $X \sim X+2\pi$ in extremal coordinates \eqref{extpat} and (equivalently) $\Phi \sim \Phi+2\pi$ in horizon coordinates \eqref{horizon-coordinates}.  We will call $\xi$ the extremal Killing field.  Its integral curves connect the two ends of the ``conformal bifurcation line'' where the future and past horizons meet at the boundary (Fig.~\ref{fig:patches}).

\section{Scalar field}\label{sec:scalar}
We consider a massive scalar field $\psi$,
\begin{align}\label{scalar}
    \left( \Box - \mu^2 \right) \psi = 0.
\end{align}
The $\text{AdS}_3$/BTZ spacetimes are not globally hyperbolic, and well-posed evolution requires specification of the behavior of the field on the boundary.  We will consider so-called ``Dirichlet'' conditions, where $\psi$ is required to vanish on the boundary.  We define an associated retarded Green function by 
\begin{align}\label{GF}
    \left( \Box - \mu^2 \right) G(x,x') = \delta_3(x,x'),
\end{align}
together with the requirements that $G$ vanish when either point approaches the boundary or if $x'$ is not in the causal past of $x$.  (Here $\delta_3$ is the covariant delta distribution, equal to $1/\sqrt{-g}$ times the coordinate delta function).  We assume $\mu^2\geq-1$ so that the Dirichlet dynamics are well-posed \cite{Breitenlohner:1982bm,Ishibashi:2004wx}.  This Green function can be used to construct the field from its initial value $\psi_0(R',\Phi')$ on the null surface $V'=0$ via the Kirchhoff representation\footnote{To derive equation \eqref{kirchhoff}, we follow the steps of Ref.~\cite{lrr-2011-7} using a volume bounded by the null surface $V=0$, the future event horizon, and the boundary.  The contribution at the boundary vanishes due to the Dirichlet condition.}
\begin{equation}\label{kirchhoff}
    \psi(V,R,\Phi) = R_H^{-1}\int d\Phi' \int_{R_H}^\infty dR' \, R' \left(\psi_0 \pd_{R'} G - G \pd_{R'}\psi_0 \right),
\end{equation}
where we have assumed that the field point is on or outside the event horizon.  
In $\text{AdS}_{3}$, the range of $\Phi'$ is unbounded, while in BTZ it must be restricted to a fiducial range such as $0 < \Phi' \leq 2\pi$. In all other respects,  Eqs.~\eqref{scalar}, \eqref{GF}, and \eqref{kirchhoff} hold equally well in $\text{AdS}_3$ and extermal BTZ.

\subsection{$\text{AdS}_3$ Green function}\label{sec:Ads3GF}
The Dirichlet retarded Green function for $\text{AdS}_3$ was obtained in closed form by \cite{Danielsson:1998wt}.  In App.~\ref{sec:GF} we review the derivation, correct some trivial errors, and obtain a new form of the result: 
\begin{align}\label{adsgreen}
    G_{\text{ret}}^{\text{AdS}_3} = \frac{ \Theta(t-t')\Theta(\Sigma)}{\pi\sqrt{|\Sigma (\Sigma-2)|}} \begin{dcases}-  \cos{2\nu\arccos{(1-\Sigma)}} \quad &\text{if} \quad \Sigma<2 \\ \sin{(2\pi \nu)} e^{-2\nu \text{arccosh}{(\Sigma -1)}} \quad &\text{if} \quad \Sigma>2 \end{dcases}.
\end{align}
Here $\nu$ is a scaling dimension for the field defined by
\begin{equation}
\nu= \frac{1}{2} \sqrt{1+\mu^2},
\end{equation} 
while $\Sigma$ is a biscalar on $\text{AdS}_3$ given by
\begin{align}
    \Sigma & = \frac{(t-t')^2 - (x-x')^2 - (z-z')^2}{2zz'} \\
    & =  1- \cos(\tau - \tau') \sec \chi \sec \chi' + \cos (\Omega - \Omega') \tan \chi \tan \chi'.
\end{align}

\begin{figure}
    \centering
     \includegraphics[width=.2\linewidth]{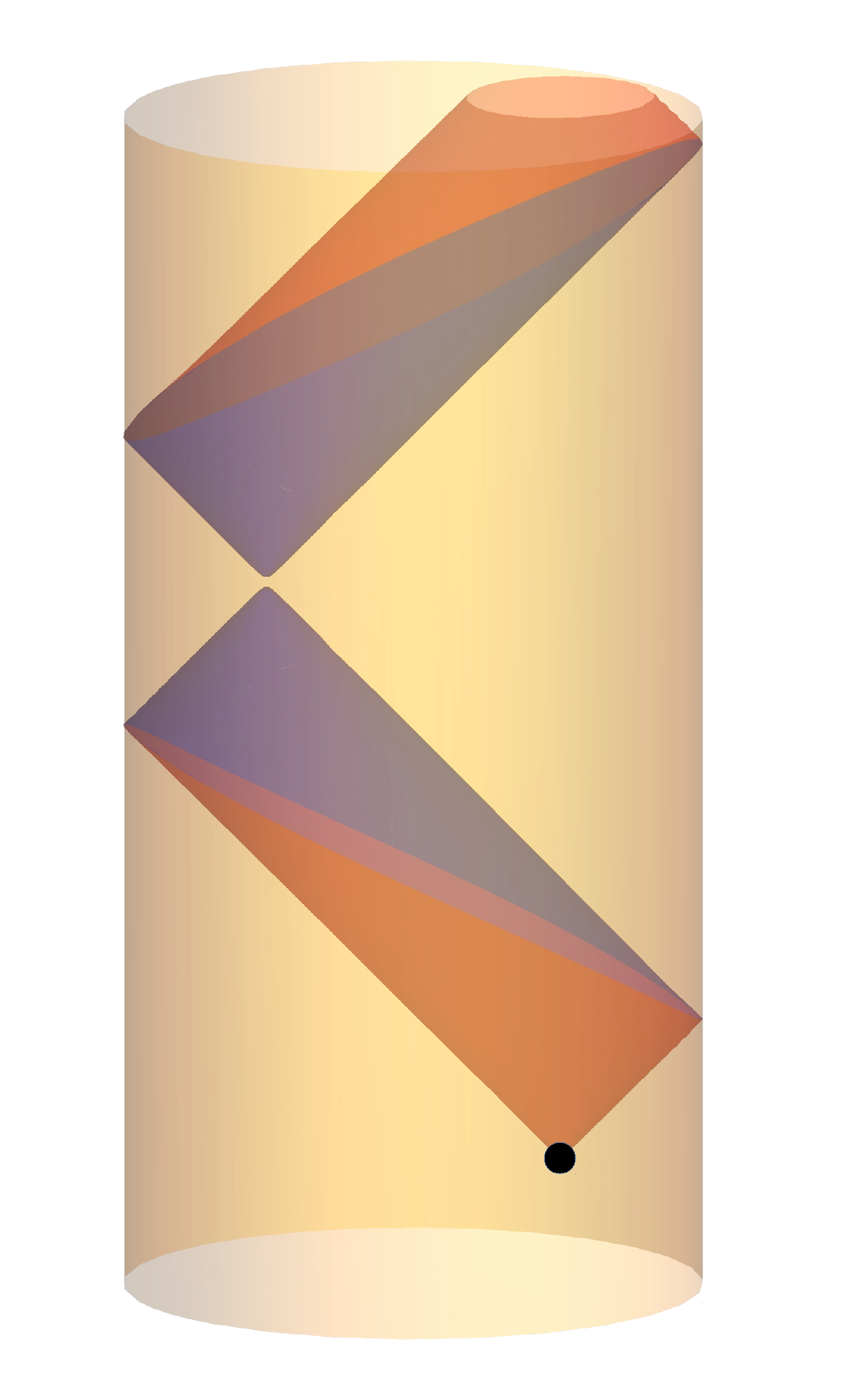}
    \includegraphics[width=.39\linewidth]{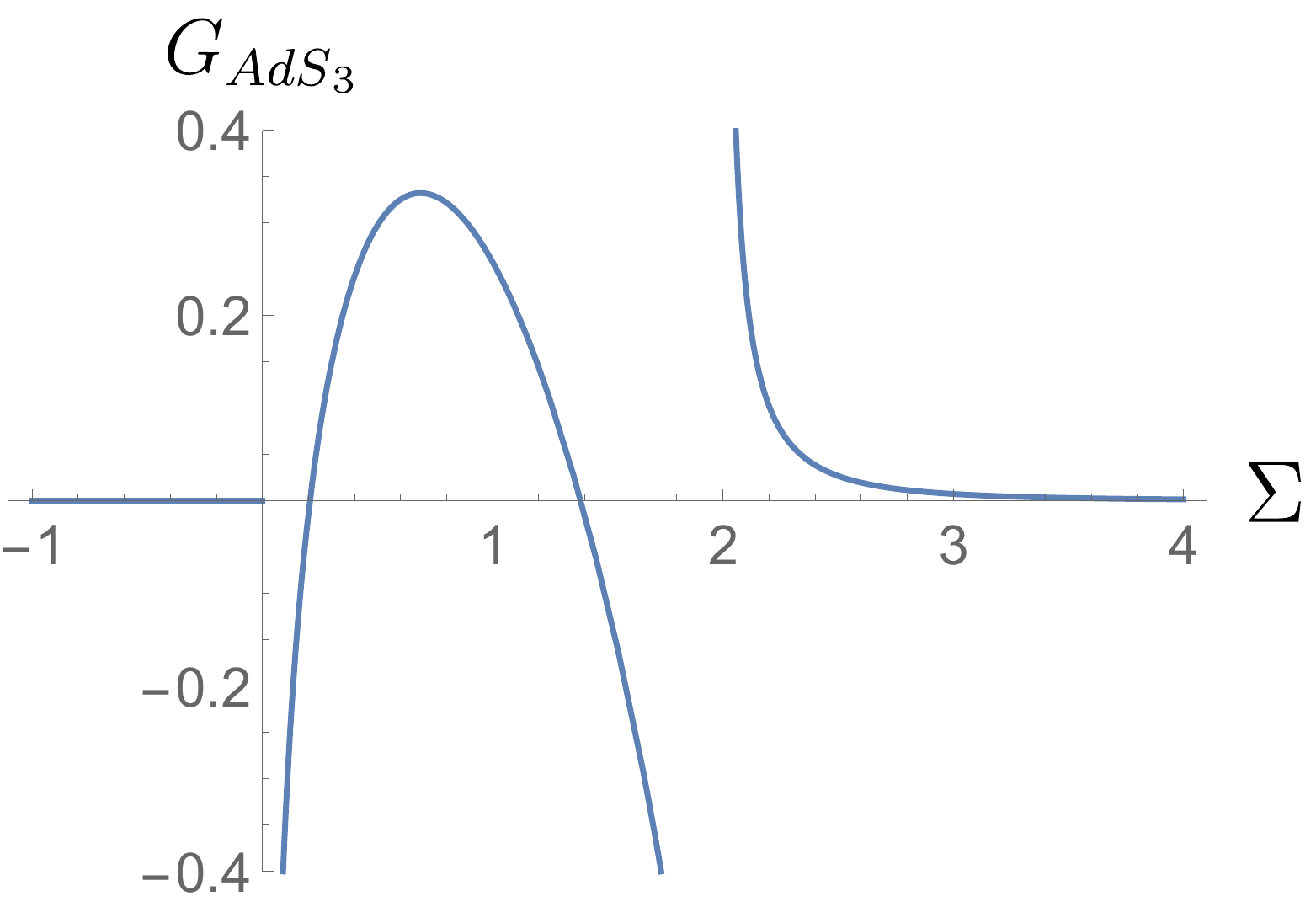}
    \includegraphics[width=.39\linewidth]{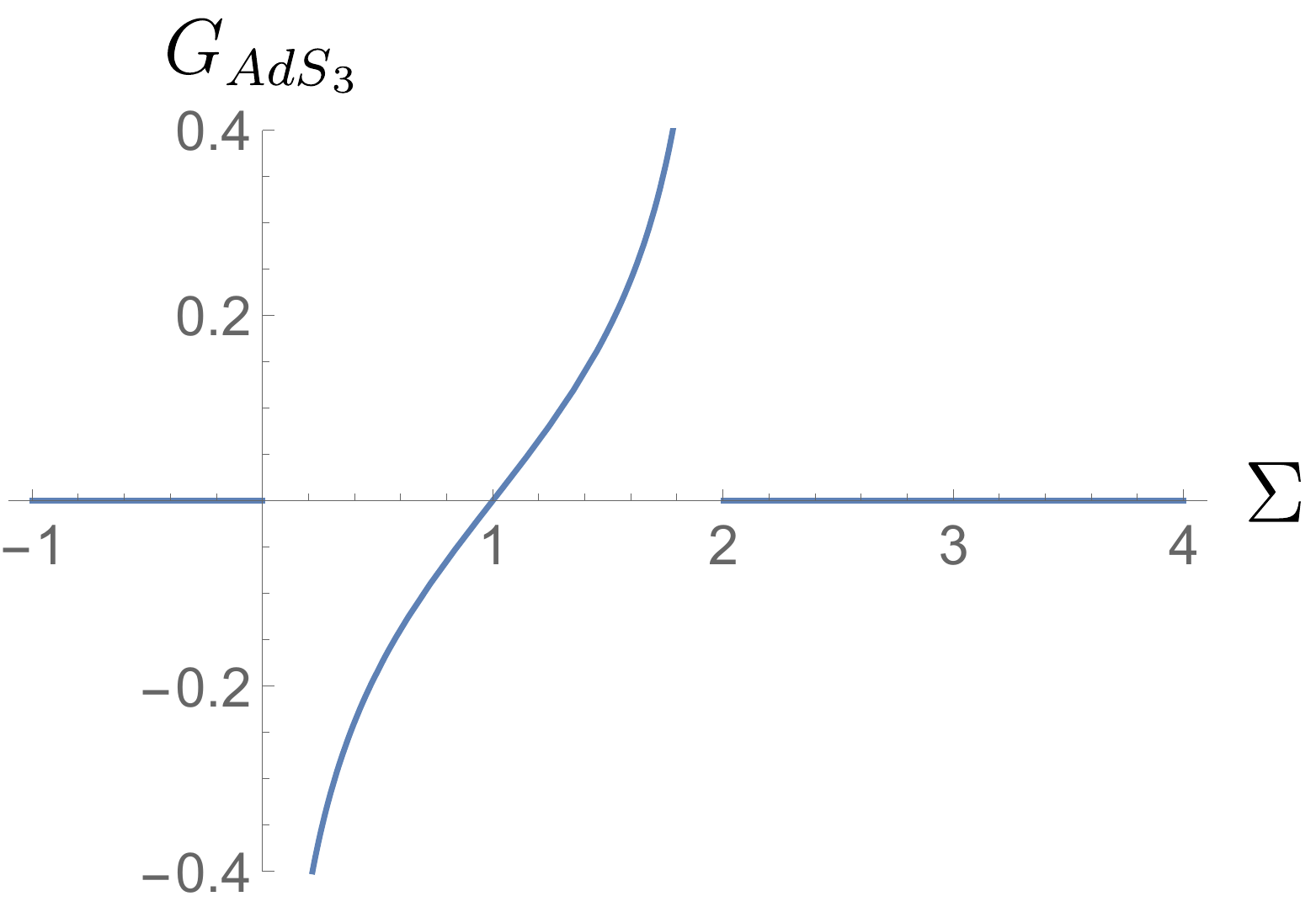}
    \caption{ Retarded propagation in $\text{AdS}_3$ with Dirichlet boundary conditions.  On the left, we show the wavefront of a point source (surface of singularity of the retarded Green function), which ``bounces'' off the boundary.  The wavefront is described by $\Sigma=0$ (orange) and $\Sigma=2$ (blue).  On the right, we show the behavior of the retarded green function as a function of $\Sigma$.  Two qualitative behaviors exist: the ``generic'' case (middle; we have chosen $\nu=1.2$) and the ``exceptional'' case (we have chosen $\nu=0.5$).  In the exceptional case, the Green function vanishes for $\Sigma>2$.}
    \label{fig:Sigma}
\end{figure}
The Green function is singular on the surfaces $\Sigma=0$ and $\Sigma=2$, whose union is the null wave front of the propagating field.  This wave front bounces off the boundary as shown in Fig.~\ref{fig:Sigma}.  In the ``exceptional'' case where $\nu$ is a half-integer, the Green function vanishes for $\Sigma>2$, so that the returning wavefront appears to ``cancel out'' the propagating field.  Although the wavefronts appear conical in the plots, they are not precisely equal to cones.  The distinction is most graphically evident when the source point is near the boundary, in which case the wavefronts collapse to Poincar\'e horizons.


We now discuss the geometric interpretation of $\Sigma$ and its role in the Green function.  In a Poincar\'e patch there is always a unique geodesic connecting two points, and $\Sigma$ is related to the world function $\sigma$ (one-half the squared geodesic distance) by \cite{Dappiaggi:2016fwc}
\begin{align}
    \Sigma & = 2 \left( 1- \cosh^2 \left( \frac{\sqrt{2\sigma}}{2}\right)\right) = -\sigma + \mathcal{O}(\sigma^2).
\end{align}
In particular, $\Sigma>0$ is timelike separation, $\Sigma<0$ is spacelike separation, and $\Sigma=0$ is null separation.  However, we have seen that $\Sigma=2$ also represents a null wavefront of the propagating field.  This is not a contradiction, since the bouncing behavior of the wave fronts is \textit{not} reflected in the behavior of geodesics, which simply continue indefinitely toward the boundary as the affine parameter increases.  Thus, at least within the Poincar\'e patch, points for which $\Sigma=2$ are indeed connected by only a single geodesic, which is timelike, despite the fact that they are also connected by a bouncing wave front of the scalar field.  The null character of the surface $\Sigma=2$ (fixing the prime point) can be seen directly by observing that it corresponds to $\Sigma=0$ for ``conjugate points'' $\tau' \to \tau'+n \pi, \ \Omega' \to \Omega'+n \pi$ for odd $n$ (see  Fig.~\ref{fig:Sigma}).  Finally, note that the wavefront preserves the inverse square root character of the singularity in the Green function inherited from the local Hadamard form.  See e.g. \cite{Kay:1996hj} and references therein for rigorous results on the global propagation of singularities.

\begin{figure}
    \centering
    \includegraphics[width=.45\linewidth]{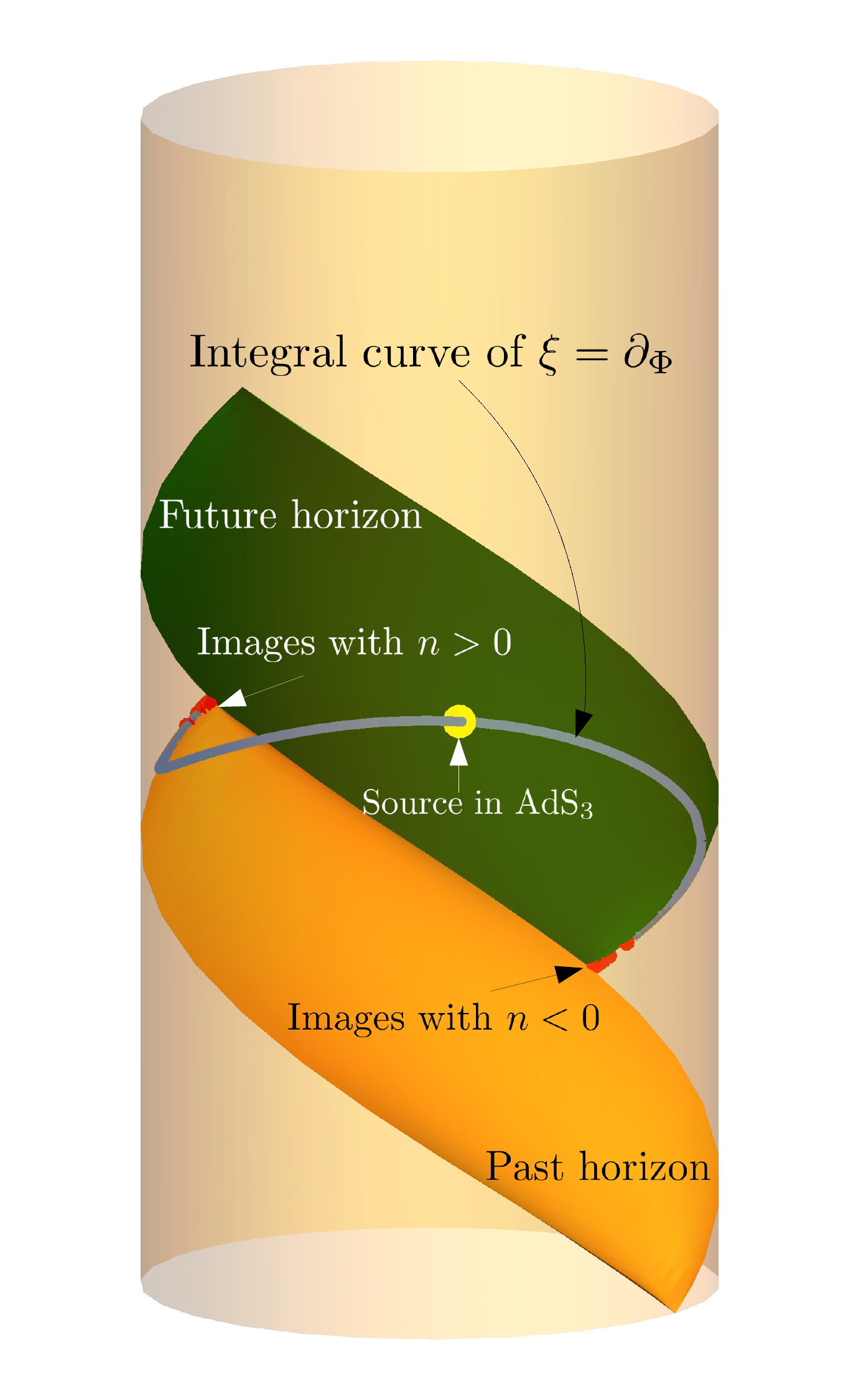}
    \centering
    \caption{Method of images for the extremal BTZ Green function.  A source point (yellow) in the fiducial portion of $\textrm{AdS}_3$ defines image points (red) appearing every $2\pi$ along the integral curve of $\xi = \pd_\Phi$.
    }
    \label{fig:images}
\end{figure}

\subsection{Extremal BTZ Green function}

As the BTZ black hole is a periodic identification of $\text{AdS}_3$, propagation in BTZ is equivalent to propagation on a portion of $\text{AdS}_3$ subject to periodic boundary conditions.  The retarded Green function can then be simply constructed from the non-periodic one by
\begin{align}\label{GBTZ}
    G_{\text{ret}}^{\text{BTZ}}(x,x') & = \sum_{n=-\infty}^{\infty}G_{\text{ret}}^{\text{AdS}_3}(x,e^{2\pi n\xi} x') 
     =\sum_{n=-\infty}^{\infty}G_{\text{ret}}^{\text{AdS}_3}|_{\Phi' \rightarrow \Phi'+2\pi n},
\end{align}
where the last equation holds only in horizon coordinates.  The notation $e^{2\pi n \xi}x'$ indicates the spacetime point obtained by flowing along the integral curves of $\xi$ for a parameter distance $2\pi n$; when horizon coordinates are used, this simply increases $\Phi'$ by $2\pi n$.  It is straightforward to check that this expression satisfies all the conditions for the Dirichlet retarded Green function, together with the additional required periodicity.  From the perspective of $\text{AdS}_3$, we place ``image sources'' at distances of $2\pi n$ along the integral curves of $\xi$ from the original source position (Fig.~\ref{fig:images}), all of which emit wave fronts.  From the perspective of BTZ, the arriving fronts are interpreted as having circled the black hole $|n|$ times.  This method of constructing the Green function is generally called the method of images; it has been used extensively in Euclidean signature, but apparently not for retarded propagation.

It will be convenient to define a separate $\Sigma_n$ for each image $n$.  Letting
\begin{align}
    \delta V = V-V', \qquad \delta \Phi_n=\Phi-\Phi'-2\pi n,
\end{align}
then (expressing in horizon coordinates) we have
\begin{align}\label{Sigman}
    \Sigma_n & = 1 - \frac{e^{R_{H}\delta \Phi_n}}{2R_{H}} \left( \frac{R'-R}{2} + R_{H} + \left(\delta V + \frac{\delta \Phi_n}{2}\right)(R'+R_{H})(R-R_{H}) \right) \nonumber \\
    & \quad + \frac{e^{-R_{H}\delta \Phi_n}}{2R_{H}} \left(  \frac{R'-R}{2} - R_{H} + \left(\delta V + \frac{\delta \Phi_n}{2}\right)(R'-R_{H})(R+R_{H})  \right).
\end{align}
The other component of the $\textrm{AdS}_3$ retarded Green function \eqref{adsgreen} is a theta function, $\Theta(t-t')$, that cuts out the past of $t'$.  We will introduce the notation
\begin{align}
    s_n=t-t'|_{\Phi' \to \Phi'+2\pi n},
\end{align}
indicating that one is to express $t$ and $t'$ in terms of horizon coordinates and then send $\Phi' \to \Phi'+2\pi n$.  Explicitly, we have
\begin{multline}\label{sn}
    s_n = \frac{1}{4(R+R_{H})(R'+R_H)} \left[ e^{2R_{H} (\Phi'+2\pi n)} (R+R_{H})(R'-R_{H}) \right.\\
    \left.- e^{2R_{H} \Phi} (R'+R_{H})(R-R_{H})  + 2R_{H}(R'-R + (R+R_{H})(R'+R_{H})(2\delta V + \delta \Phi_{n})) \right].
\end{multline}
Putting everything together, the BTZ Dirichlet retarded Green function is written
\begin{align}\label{Gfinal}
    G_{\text{ret}}^{\text{AdS}_3} = \sum_{n=-\infty}^\infty\Bigg(\frac{ \Theta(s_n)\Theta(\Sigma_n)}{\pi\sqrt{|\Sigma_n (\Sigma_n-2)|}}  \begin{dcases}-  \cos{2\nu\arccos{(1-\Sigma_n)}} \ &\text{if} \quad \Sigma_n<2 \\ \sin{(2\pi \nu)} e^{-2\nu \text{arccosh}{(\Sigma_n -1)}} \ &\text{if} \quad \Sigma_n>2 
   \end{dcases}  \Bigg),
\end{align}
where $\Sigma_n$ and $s_n$ are given in Eqs.~\eqref{Sigman} and \eqref{sn}, respectively.  The non-uniform behavior of $\Sigma_n$ in the limits $V \to \infty$, $n \to \pm \infty$, $R' \to R_H$ and $R \to R_H$ is key to the phenomena discussed below.

\section{Late-time Decay}\label{sec:late}

We now use the explicit retarded Green function to investigate late-time decay of field perturbations.  We will discuss decay in null directions, varying $V$ while fixing $R$ and $\Phi$.  When $R=R_H$ this corresponds to decay along a horizon generator $\Phi$, measured by affine parameter $V$.  When $R>R_H$ the null ray remains outside the horizon.


\subsection{$\textrm{AdS}_3$}

We begin with the case of $\textrm{AdS}_3$.  The $\textrm{AdS}_3$ Green function is simply the $n=0$ term in the sum \eqref{Gfinal}.  At large $\delta V$ we have
\begin{align}\label{SigmaHorLate}
    s_0 & \sim R_{H}\delta V, \qquad \qquad \qquad \quad \delta V \to \infty, \\
    \Sigma_0 & \sim C_0(R,R',\delta \Phi_0) \ \! \delta V, \qquad \ \ \! \delta V \to \infty,
\end{align}
with
\begin{align}
    C_0=\frac{1}{2 R_H}\Big[e^{-R_{H}\delta \Phi_0} (R'-R_{H})(R+R_{H})  - e^{R_{H}\delta \Phi_0}  (R'+R_{H})(R-R_{H})\Big],
\end{align}
where we exclude the measure-zero case $C_0=0$.  If $C_0<0$ then $\Sigma_0$ is negative at late times and the Green function vanishes: the points of fixed $R$ and $\Phi$ with large $V$ are out of causal contact with the source position $R',\Phi',V'$.  When $C_0>0$, the Green function at late times is
\begin{align}
G \sim \frac{4^{-\nu} C_0}{\pi} \sin(2\pi \nu) (\delta V)^{-(2\nu+1)}, \qquad \delta V \to \infty.
\end{align}
If $\nu$ is a half-integer then this expression vanishes.  In fact the Green function \eqref{Gfinal} vanishes identically in this case,
\begin{align}
G = 0, \qquad \delta V > V_0,
\end{align}
where $V_0$ is some constant. By the Kirchhoff integral \eqref{kirchhoff}, the field sourced by generic initial data will behave the same way,
\begin{align}
\psi & \sim C(R,\Phi) V^{-2h}, \qquad V \to \infty \qquad (2h\notin\mathbb{Z}^+)\\
\psi & = 0, \qquad \qquad \quad \ \qquad V>V_0 \ \! \qquad (2h\in\mathbb{Z}^+).
\end{align}
where $C$ is some function determined by the initial data, and we introduced
\begin{align}\label{eq:h}
    h & = \frac{1}{2} + \nu 
\end{align}
in order to be consistent with previous work \cite{Gralla:2018xzo, Gralla:2017lto,Gralla:2016sxp}. Notice that $h\geq1/2$ given our assumption on the mass $\mu^2\geq-1$.  We will refer to the case $2h\notin \mathbb{Z}^+$ as ``generic'' and the case $2h\in \mathbb{Z}^+$ as ``exceptional''.  This exceptional case includes the case $h \in \mathbb{Z}^+$ previously called ``discrete''.

\subsection{BTZ}

For a BTZ black hole, we must consider the infinite sum \eqref{Gfinal}. Each term in the sum is an $\textrm{AdS}_3$ Green function with a different source point, so each term behaves like $\sin(2\pi \nu)V^{-2h}$ at late times as above.  However, this does not guarantee that the full sum shares this behavior, and indeed we will see that it does not.  To examine the full sum, it is helpful to rewrite \eqref{Sigman} as
\begin{align}
    \Sigma_n = 1+ e^{2 \pi n R_H}(A_+ + B_+ (\delta V - \pi n)) -  e^{-2 \pi n R_H} (A_- + B_- (\delta V - \pi n)),
\end{align}
with
\begin{subequations}
\begin{align}
A_\pm & = \frac{e^{\mp R_H (\Phi-\Phi')}}{2 R_H} \left( \frac{R'-R}{2} \mp R_H \right)\\
    B_\pm& =\frac{e^{\mp R_H (\Phi-\Phi')}}{2 R_H} (R'\mp R_{H})(R\pm R_{H}).
\end{align}
\end{subequations}
To explore the deviation of the full sum from the behavior of an individual term we consider the regime of late times and large image numbers, i.e. large $\delta V$ and large $|n|$.  We restrict to the generic case $2h \notin \mathbb{Z}^+$.

\begin{figure}
    \centering
    \begin{minipage}{.48\textwidth}
        \includegraphics[width=.98\linewidth]{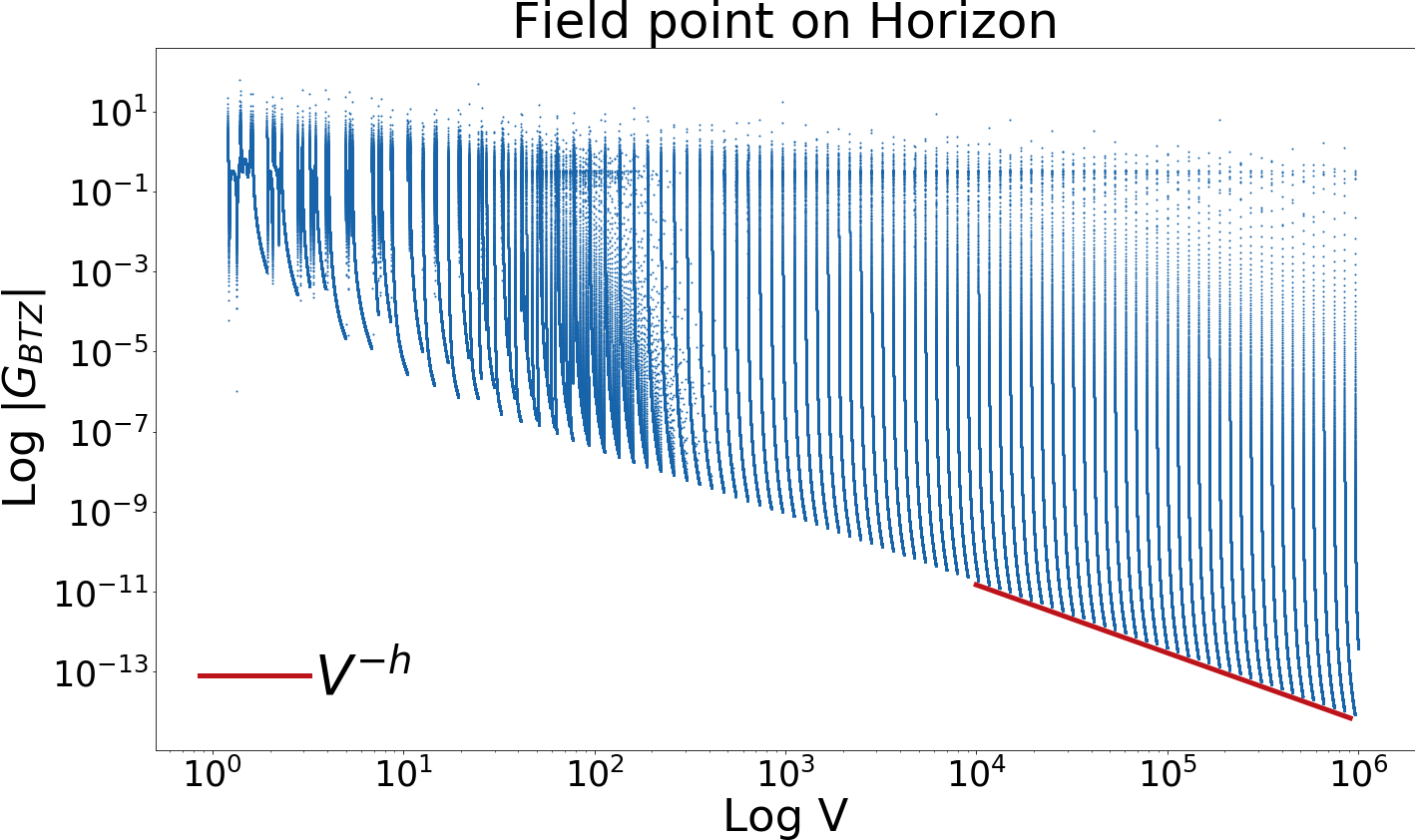}
    \end{minipage}
    \begin{minipage}{.48\textwidth}
        \includegraphics[width=.98\linewidth]{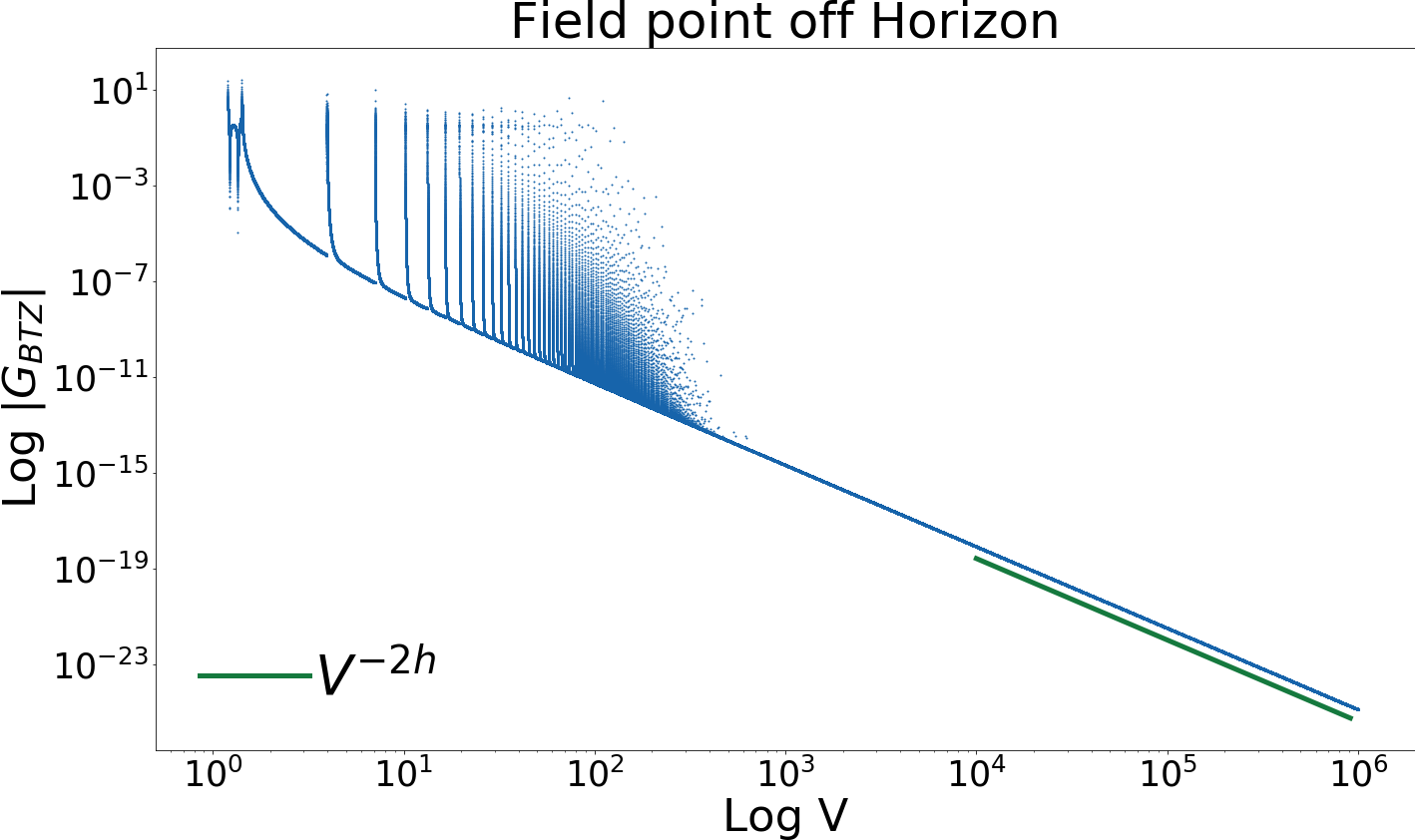}
    \end{minipage}
    \caption{The BTZ retarded Green function with field point off the horizon, sampled uniformly in $\log V$. The field decays at different rates depending on whether the field point is on the horizon (left plot, $V^{-h}$) or off the horizon (right plot $V^{-2h}$).  The slower decay is due to spikes from wavefronts that orbit many times near the black hole and cross the horizon at arbitrarily late times. Spikes also arrive at arbitrarily late times from wavefronts that spend their time mainly far away from the black hole, but they are exponentially narrow and are not visible in the plots at late times (and do not affect the decay envelope). In these plots we have chosen $R_H=0.01$ and $\nu=1.2$ such that $h=1.7$, with $\delta \Phi_0=0$ and $R'=10$; in the left plot we have $R=R_H=0.01$, while in the right plot we have $R=1>R_H$.}
    \label{fig:decay}
\end{figure}

\subsubsection{Both points outside the horizon}

 Suppose now that both the source and field point are outside the horizon, $R>R_H$ and $R'>R_H$.  Then $B_\pm$ are both non-zero, so in the regime of large $\delta V$ and $|n|$ we have
\begin{align}\label{Sigmanoffoff}
    \Sigma_n \approx \begin{cases} B_+ e^{2\pi n R_H} (\delta V - \pi n), &  n>0 \\ - B_- e^{2\pi |n| R_H} (\delta V + \pi |n|), &  n<0\end{cases}, \qquad \delta V \gg 1, \ |n| \gg 1.
\end{align}
The contribution of the $n^{\rm th}$ term to the Green function is important when $\Sigma_n$ is positive and of order unity (see Fig.~\ref{fig:Sigma}), with the initial spike occurring when $\Sigma_n=0$.  Since $B_\pm$ are positive, we see that the negative $n$ terms do not contribute ($\Sigma_n$ is negative), while the positive $n$ terms are important for times of order the image number ($\delta V\sim n$).  The importance of the contribution can be assessed from the derivative
\begin{align}\label{derivplus}
    \frac{\pd \Sigma_n}{\pd V} \approx B_+ e^{2\pi n R_H}, \qquad (n>0).
\end{align}
Since this derivative is exponentially large, the width (in $V$) of the region of importance is exponentially suppressed with image number.  This suggests that the large-$n$ terms are not very important, and (aside from brief exponentially narrowing spikes) the full behavior of the sum will share the falloff $V^{-2h}$ of each individual term.  Plotting the Green function numerically supports this conclusion  (Fig.~\ref{fig:decay} left).

The behavior of the field $\psi$ follows from that of the Green function via the Kirchhoff integral \eqref{kirchhoff}.  It is clear from the discussion above that the Green function decays as $V^{-2h}$ apart from exponentially narrowing spikes that may be safely excluded from the integral.  This late-time approximation is uniformly valid over any region of $R'$ outside (but not including!) the event horizon, so the field will similarly decay as $V^{-2h}$ provided the initial data is confined outside the horizon.  The spikes in the Green function will be smoothed out by the integral into finite oscillations about this decay.  That is, we claim that the generic outside-horizon, late-time behavior of fields sourced by initial data outside the horizon is
\begin{align}\label{psioff}
    \psi \sim \psi_0 V^{-2h}, \qquad R>R_H, V \to \infty
\end{align}
where $\psi_0$ is an arbitrary function of $R,\Phi$ and an $O(1)$ function of $V$.

\subsubsection{Field point on the horizon and source point outside}

If the field point is now on the horizon $(R=R_H)$ while the source is still outside ($R'>R_H$), then $B_-=0$ and at large $\delta V$ and $|n|$ we have
\begin{align}\label{Sigmanonoff}
    \Sigma_n \approx \begin{cases} B_+ e^{2\pi n R_H} (\delta V - \pi n), &  n>0 \\ B_+e^{-2\pi |n| R_H}(\delta V - \pi n) - A_- e^{2\pi |n| R_H}, &  n<0\end{cases}, \qquad \delta V \gg 1, |n| \gg 1.
\end{align}
which may be compared with Eq.~\eqref{Sigmanoffoff}.  The formula for the positive $n$ terms is identical, and these terms again contribute at $\delta V \sim n$.  However, now the negative $n$ terms contribute as well, starting at even later times $\delta V \sim e^{4\pi |n|R_H}$ (where $\Sigma_n$ becomes positive).  The derivatives in these regimes are given by 
\begin{align}
    n>0: \qquad \frac{\pd \Sigma_n}{\pd V} & \approx B_+ e^{2\pi |n| R_H},  \qquad  (n \to \infty, \quad \delta V \gtrsim \pi n), \label{linearvgap}\\
    n<0: \qquad \frac{\pd \Sigma_n}{\pd V}& \approx B_+ e^{-2\pi |n| R_H} \qquad (n \to -\infty, \quad \delta V \gtrsim \frac{A_-}{B_+}e^{4\pi |n|R_H}),\label{logvgap}
\end{align}
which may be compared to Eq.~\eqref{derivplus}.  The positive $n$ terms again have a large derivative, indicating an exponentially narrow contribution to the Green function.  However, the newly contributing negative $n$ terms have a small derivative, indicating an exponentially \textit{wide} spike in the Green function as a function of $V$.  Thus the negative $n$ spikes should be very important at late times, and there is no reason to expect that the sum has the same falloff as an individual spike. Plotting the Green function (Fig.~\ref{fig:decay} right) confirms the exponentially wide spikes and shows that the actual decay is $V^{-h}$.  

We can understand this rate through the following heuristic argument.  Let us label the exponentially wide spikes by $m=-n$, so that $m$ is a positive integer.  As discussed above, the $m{}^{\rm th}$ spike arrives at a time of order $\delta V \sim e^{4\pi m R_H}$.  (In this discussion we drop the $m$-independent constants $A_-$ and $B_+$.)  At the precise time of arrival we have $\Sigma_m=0$, but shortly thereafter (including the majority of the time before the arrival of the next spike) we have $\Sigma_m \sim e^{2\pi m R_H} \gg 1$.  Thus we may approximate each image's contribution $G_m$ to the green function as its large-$\Sigma$ behavior $G_m \sim \Sigma_m^{-2h}$ [Eqs.~\eqref{Gadslate} and \eqref{eq:h}].  Focusing on the period of order $\delta V \sim e^{4 \pi m R_H}$ before the next spike has arrived, we may equivalently express this in terms of $m$ or $\delta V$ as
\begin{align}
    G_m \sim \Sigma_m^{-2h} \sim e^{-4 \pi h m R_H} \sim (\delta V)^{-h}, \qquad \delta V \sim e^{4\pi m R_H},
\end{align}
recalling that $m$-independent constants are dropped.  This shows the $(\delta V)^{-h}$ decay visible in Fig.~\ref{fig:decay}.  In essence, each image decays as $C_m (\delta V)^{-2h}$, but the effective amplitudes $C_m$ grow as $(\delta V)^h$, resulting in an overall $(\delta V)^{-h}$ decay.

By similar arguments made for Eq.~\eqref{psioff} above, we expect that the generic behavior of fields sourced by initial data outside the horizon is thus
\begin{align}
    \psi \sim \psi_H V^{-h}, \qquad R=R_{H},  V \to \infty
\end{align}
where $\psi_H$ is an arbitrary function of $\Phi$ and an $O(1)$ function of $V$.

For later comparison it will be helpful to have a more precise expression for the time of arrival of the $n^{\textrm{th}}$ spike.  This is obtained by letting $\delta V \gg n$ and $\Sigma_n=0$ for $\delta V$, giving
\begin{align}
    \delta V & \sim \pi n,  \!\!\! &n \to\infty, \label{deltaVlargen} \\
    \delta V & \sim  \frac{e^{2 R_{H} (\Phi-\Phi')}}{4R_H}\frac{R'+R_{H}}{R'-R_{H}} e^{4 \pi |n| R_{H}}, &n \to -\infty. \label{deltaVlargenegativen}
\end{align}

\begin{figure}
    \centering
    \includegraphics[width=.50\linewidth]{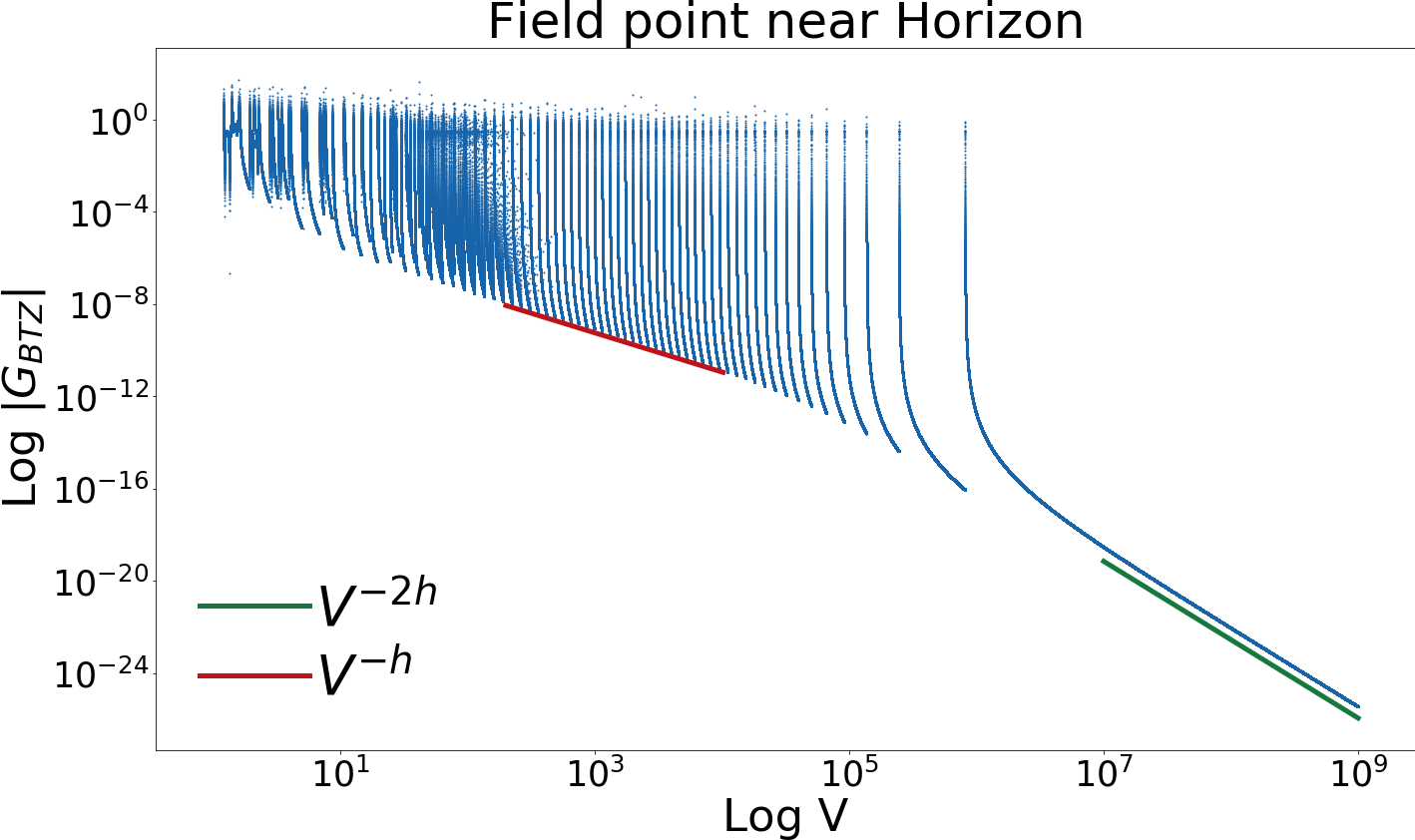}
    \caption{Illustration of transient $V^{-h}$ decay before ultimate $V^{-2h}$ decay for field points near the horizon.  The parameters are the same as Fig.~\ref{fig:decay} except we take $(R-R_H)/R_H=10^{-3}$.}
    \label{fig:transient}
\end{figure}

If the field point is \textit{near} the horizon but not exactly on it, one expects a transient period of $V^{-h}$ decay followed by final decay of $V^{-2h}$.  Plotting the Green function confirms this expectation (Fig.~\ref{fig:transient}).  The field $\psi$ sourced by initial data outside the horizon will similarly show this transition from $V^{-h}$ to $V^{-2h}$ when evaluated near the horizon.

\subsubsection{Both points on the horizon}

If both the source and field point are on the horizon ($R=R'=R_H$), then we have the simple expression
\begin{equation}
    \Sigma_n = 1-\cosh \left( R_{H} (\delta \Phi_0 - 2\pi n )\right).
\end{equation}
In particular, the Green function is actually independent of $V$ (apart from the causal factor $\Theta(s_n)$).  Of course, for a source point at at any small distance off the horizon the Green function does decay, with the rate determined by whether the field point is on or off the horizon.  That is, the late-time behavior of the Green function is highly non-uniform as the points approach the horizon.  Determining the decay rate of fields with initial data that extends to the horizon would require a careful estimate of the Kirchhoff integral \eqref{kirchhoff} using the full behavior of the Green function near the horizon.  We leave this to future work.



\section{Null geodesics}\label{sec:geodesics}

We now study the null geodesics of the BTZ spacetime in order to understand the behavior of the Green function that gives rise to the instability.  Let $p^\mu$ denote the four-momentum of the null geodesic, where $p^V>0$ is our time orientation.  Using the null condition and the two Killing fields $\zeta=\pd_V$ (horizon Killing field) and $\xi=\pd_\Phi$ (axial Killing field), we have three conserved quantities,
\begin{align}\label{conserved quantities}
    g_{\mu \nu} \xi^\mu p^\nu = L, \qquad g_{\mu \nu}\zeta^\mu p^\nu = L-E, \qquad g_{\mu \nu} p^\mu p^\nu = 0.
\end{align}
The quantity $L$ is interpreted as the angular momentum, while $E$ is the energy according to the ``static'' Killing field $\pd_T=\zeta-\xi=\pd_V-\pd_\Phi$.  Note that $E$ can be negative for trajectories inside the ``ergoregion'' $R<\sqrt{2} R_H$, where $\pd_T$ is spacelike \cite{Carlip:1995qv}.

The special case $E=L=0$ corresponds to the horizon generators,
\begin{align}
    E=L=0: \qquad p \propto \pd_V, \quad R=R_H \qquad (\textrm{horizon generators}).
\end{align}
Henceforth we will consider the region outside and including the horizon,
\begin{align}\label{outhor}
    R \geq R_H >0.
\end{align}
We will further assume $E \neq 0$ and introduce 
\begin{align}
    b=\frac{L}{E}, \qquad k^\mu = \frac{d x^\mu}{d\lambda} = \frac{p^\mu}{|E|},
\end{align}
which introduces an energy-rescaled affine parameter $\lambda$ that increases toward the future.  The case $E=0, L \neq 0$ can be handled straightforwardly by the limit $b \to \infty$, and we shall see that the case $E=L=0$ is recovered (more subtly) by $b \to 1$.  Solving Eqs.~\eqref{conserved quantities} assuming $R \geq R_H>0$ and $E \neq 0$, we find
\begin{subequations}
\begin{align}\label{geodesic}
    k^V &= \frac{s(R^2-bR_{H}^2) +R^2 k^{R}}{(R^2 - R_{H}^2)^2},\\
    k^R &= \pm \frac{\sqrt{(1-b)(-2bR_{H}^2 + (1+b) R^2)}}{R}, \\
    k^{\Phi} &= -\frac{s(1-b) + k^R}{(R^2 - R_{H}^2)},
\end{align}
\end{subequations}
where $s=\textrm{sgn}(E)$.

The requirement that $k^R$ be real imposes an allowed region of $R$ for each value of $b$. Taking into account the assumption \eqref{outhor}, we find
\begin{subequations}\label{ranges}
\begin{align}
    R_{H} \leq &R \leq R_{\textrm{turn}},  \qquad |b|>1 \label{rangeb1}\\
    R_{H} \leq &R < \infty,  \ \quad \qquad |b|< 1 ,
\end{align}
\end{subequations}
with turning point radius (for $|b|>1$ only)
\begin{align}
    R_{\textrm{turn}} = \sqrt{\frac{2b}{b+1}} R_H.
\end{align}
Thus there are three kind of trajectories outside the horizon: transits from the (white hole) horizon to the boundary ($|b|<1$), transits from the boundary to the (black hole) horizon (also with $|b|<1$), and transits from the white hole to the black hole ($|b|>1$).  There are no trajectories that begin at the boundary, turn, and end at the boundary.

The sign $s=\textrm{sgn}(E)$ is fixed by the time orientation $k^V>0$ (equivalently $p^V>0$) and may be expressed directly in terms of $b$, as follows.  Since $|-b R_{H}^2 + R^2| \geq |R^2 k^R|$ for $R \geq R_{H}$, we have $\textrm{sgn}(k^V)=s \times  \textrm{sgn}(R^2-bR_H^2)$ so that $s=\textrm{sgn}(R^2-bR_H^2)$ to ensure $k^V>0$.  Given the allowed ranges \eqref{ranges} of $R$, we find that
\begin{align}
 s=\textrm{sgn}(1-b).
\end{align}
In the case $b=1$ the sign is indeterminate, with the limits $b \to 1^\pm$ having qualitatively different behavior.

\begin{figure}
    \centering
    \begin{minipage}{.27\textwidth}
        \includegraphics[width=.98\linewidth]{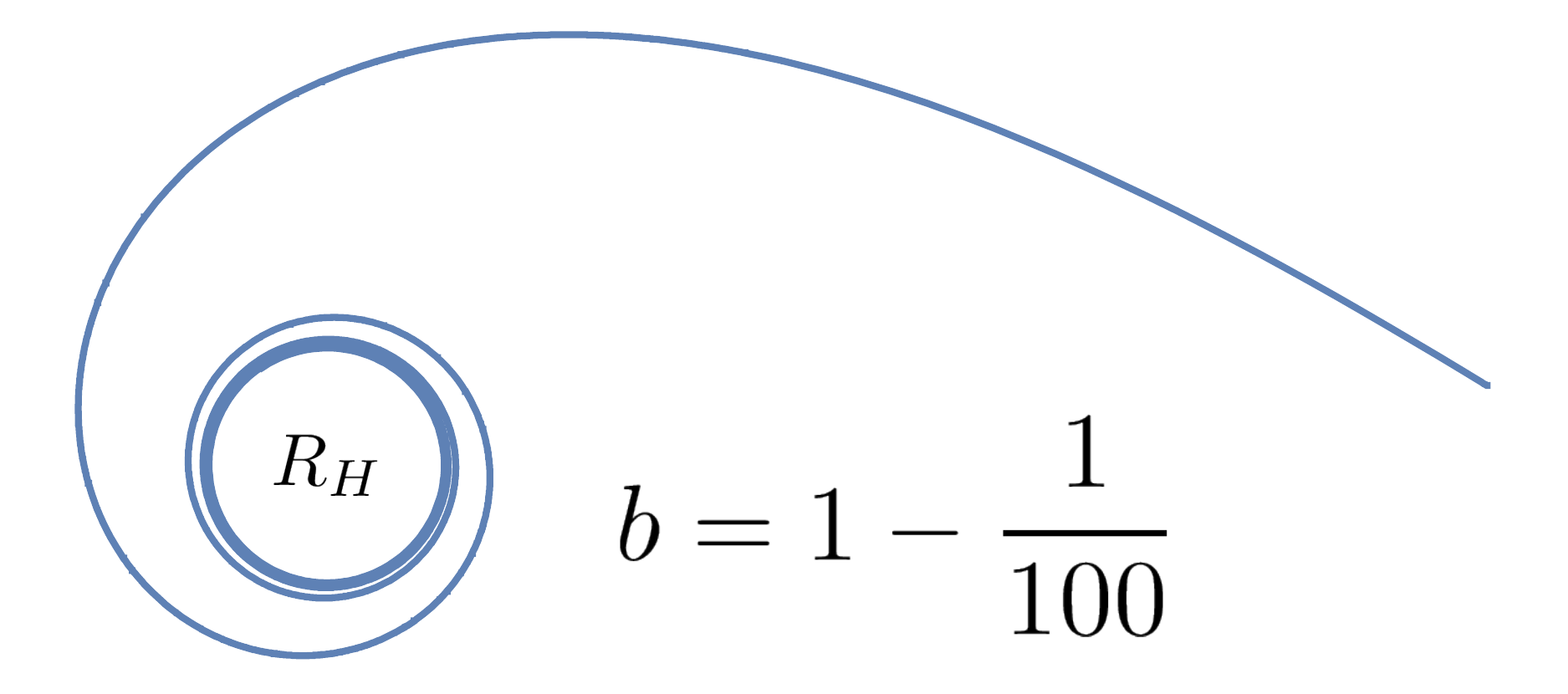}
    \end{minipage}
    \begin{minipage}{.29\textwidth}
        \includegraphics[width=.98\linewidth]{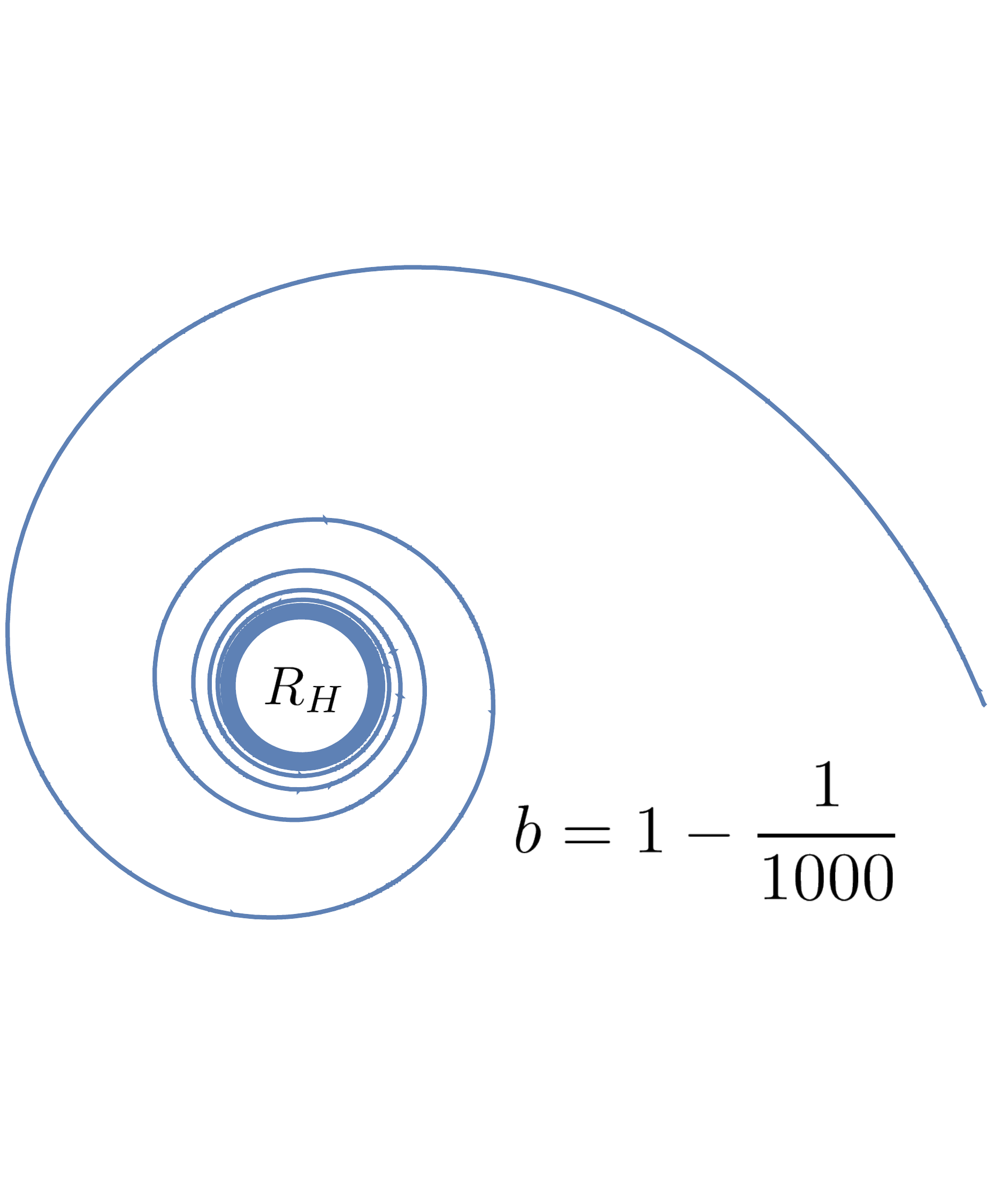}
    \end{minipage}
    \begin{minipage}{.42\textwidth}
        \includegraphics[width=.98\linewidth]{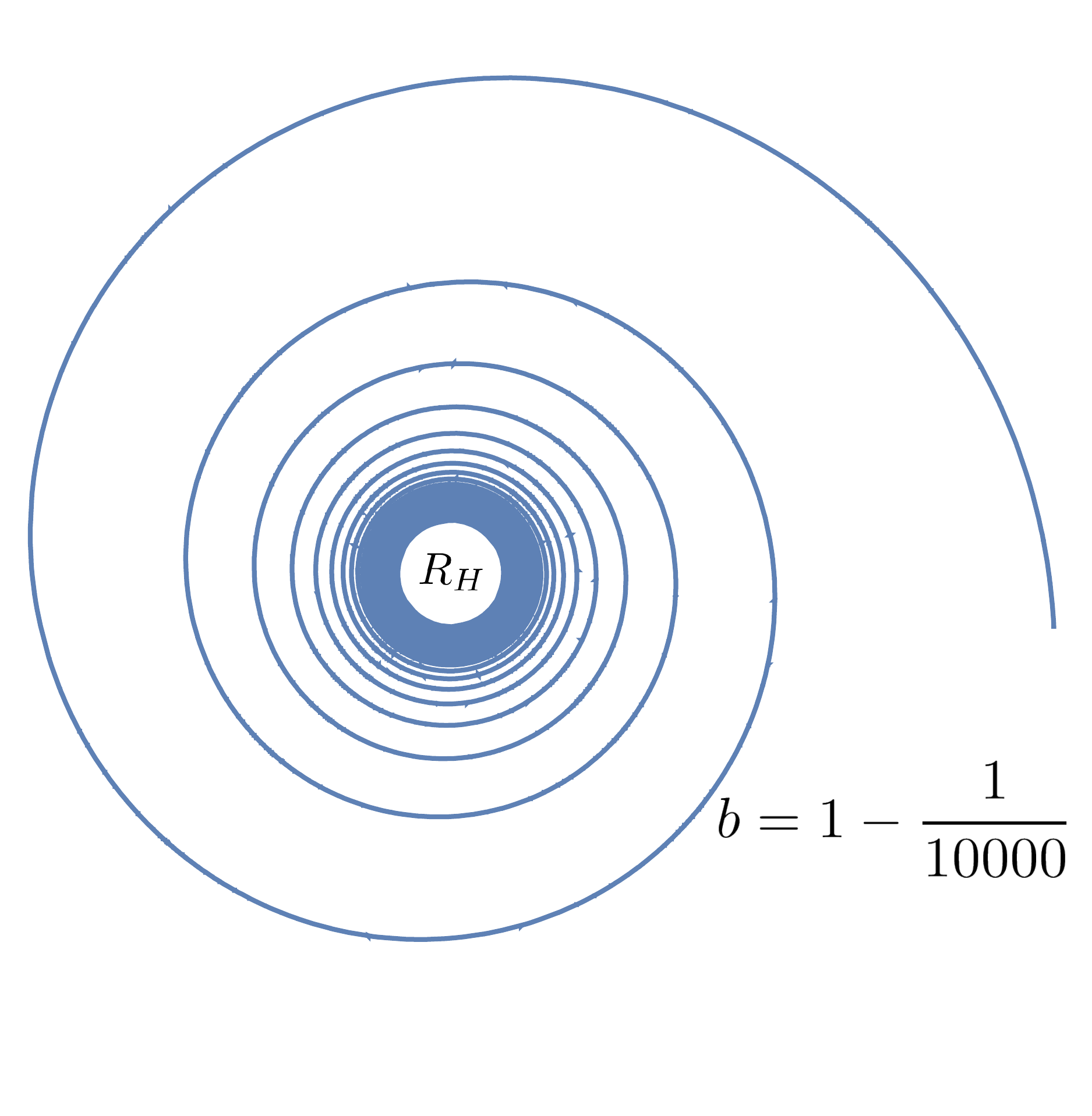}
    \end{minipage}
    \caption{Trajectories of the $b \to 1^-$ null geodesics that account for the Aretakis instability.  The angle $\Phi$ used in the text co-rotates with the black hole, so we instead use $\varphi=\Phi+V$ for these plots, treating $(R,\varphi)$ as polar coordinates in the plane.  The curves begin at a fixed $R=R_0>R_H$ and are terminated when the particle enters the horizon at $R=R_H$.  The number of orbits scales as $1/(1-b)$. }
    \label{fig:plot_geodesics}
\end{figure}



It is possible to solve Eqs.~\eqref{geodesic} entirely for the trajectory $x^\mu(\lambda)$.  Choosing the integration constants for simplicity, we find 
\begin{subequations}\label{nicesolution}
\begin{eqnarray}
     V(\lambda) &=& \frac{1}{2}\left(\frac{1+b}{1-b}\right)\frac{s(1+b) \lambda \pm  \text{ sgn}((1-b^2)\lambda) R(\lambda)}{(R_{H}^2 -(1+b)^2 \lambda^2)} - \frac{1}{2}\Phi_{\pm}(\lambda)\\
    R(\lambda) &=& \sqrt{\frac{2bR_{H}^2 +(1-b)(1+b)^2 \lambda^2}{1+b}}\\
    \Phi (\lambda) &=& \mp \frac{1}{R_{H}} \log \left|   \sqrt{\left|\frac{1+b}{1-b}\right|} \frac{ R(\lambda) -\text{sgn}((1-b^2)\lambda) R_{H}}{R_{H} \pm s(1+b) \lambda} \right| ,
\end{eqnarray}
\end{subequations}
where $\pm$ is the sign of $dR/d\lambda$ (i.e. + when outgoing and - when ingoing).  The three integration constants may be restored by shifting $V$, $\Phi$, and $\lambda$ by (separate) constant values.  With the above choices, the range of $\lambda$ is
\begin{subequations}
\begin{align}
    -\infty  &< \lambda \leq R_H/(1+b) \qquad |b|<1, \textrm{ingoing} \\
    R_H/(1+b)  &< \lambda < \infty \qquad \qquad \qquad |b|<1, \textrm{outgoing} \\
    - R_H/(1+b)  &< \lambda \leq R_H/(1+b), \qquad \! \! |b|>1, \textrm{turning at $\lambda=0$}
\end{align}
\end{subequations}
As $\lambda \to \pm R_H/(1+b)$ the particle approaches the horizon $R \to R_H$, whereas as $\lambda \to \pm \infty$ the particle approaches the boundary $R \to \infty$.  

We now fix a starting radius $R_{0}>R_H$ and compute the total change in time ($\Delta V=V-V_0$) and angle ($\Delta \Phi=\Phi-\Phi_0$) before the particle enters the horizon.  These quantities also depends on the initial radial direction (ingoing or outgoing) when $|b|>1$, since the particle will encounter a turning point if initially directed outwards.  Considering only trajectories that end at the horizon, We will label initially ingoing trajectories (present for all $b$) with ``nt'' (for ``no turning point'') and initially outgoing trajectories (present only for $|b|>1$) with ``t'' for ``has a turning point''.  Eqs.~\eqref{nicesolution} show that the lapse in time and angle is
\begin{subequations}\label{coolsolution}
\begin{eqnarray}
    \Delta V_{\text{nt}} &=& \frac{b}{2R_{H}(1-b)} -\frac{R_a -\text{sgn}(1-b) R_0}{2(R_{0}^2 - R_{H}^2)} -\frac{1}{2}\Delta \Phi_{\text{nt}} ,\qquad \forall b \label{cool1} \\
    \Delta \Phi_{\text{nt}}&=& -\frac{\text{sgn}(1-b)}{R_{H}}\log \left| \frac{R_0 + R_{H}}{R_{H}+\text{sgn}(1-b)R_a}\right| , \qquad \qquad \ \,  \forall b \label{cool2}\\
    \Delta V_{\text{t}} &=& \frac{b}{2R_{H}(1-b)} -\frac{-R_a + \text{sgn}(b) R_0}{2(R_{0}^2 - R_{H}^2)}-\frac{1}{2}\Delta \Phi_{\text{t}},\quad \ \qquad |b|>1 \label{cool3}\\
    \Delta \Phi_{\text{t}}&=&-\frac{\text{sgn}(b)}{R_{H}}\log \left| \left(\frac{1+b}{1-b}\right)\frac{R_0-R_{H}}{R_{H}-\text{sgn}(b)R_a}\right|, \qquad \quad \ \ \  |b|>1\label{cool4}
\end{eqnarray}
\end{subequations}
where
\begin{equation}
    R_a =\sqrt{\frac{R_{0}^2(1+b) -2b R_{H}^2}{1-b}}.
\end{equation}
If the quantity $R_a$ is not real, then there are no geodesics linking $R_0$ to the horizon.  This occurs when $|b|<1$ and the associated turning point is smaller than $R_0$. 

Eqs.~\eqref{coolsolution} display divergences as $b \to 1$, corresponding to geodesics that circle the horizon many times before falling in.  However, as $b \to 1^+$ (i.e. from above) the turning point moves to the horizon and there are no longer any trajectories linking $R_0>R_H$ to the horizon (also seen by $R_a$ becoming imaginary).  These $b \to 1^+$ trajectories emerge from the past horizon and orbit arbitrarily many times near the horizon radius before falling in, and are not relevant to the late-time behavior of fields from initial data confined outside the horizon, for which $R_0$ can be considered fixed.\footnote{If we scale $R_0\to R_H$ at the same rate as $b \to 1^+$ then we can recover the geodesics that turn arbitrarily close to the horizon.  These are likely relevant to the case where initial data extends to the horizon.}  For the relevant trajectories $b \to 1^-$ (i.e. from below) that originate from some fixed $R_0>R_H$, we may expand to obtain (dropping the label ``nt'')
\begin{subequations}
\begin{align}
    \Delta V &= \frac{1}{2R_{H}(1-b)} - \frac{1}{\sqrt{1-b}}\frac{1}{\sqrt{2 (R_{0}^2 - R_{H}^2)}} +\frac{1}{4R_{H}} \log \left|\frac{1-b}{2}\left( \frac{R_{0} + R_{H}}{R_{0}-R_{H}} \right)\right| + \mathcal{O} \left( 1\right)\\
    \Delta \Phi &=-\frac{1}{2R_{H}} \log \left|\frac{1-b}{2}\left( \frac{R_0 + R_{H}}{R_0-R_{H}} \right)\right| + \mathcal{O} \left( 1\right).
\end{align}
\end{subequations}
To leading order we thus have
\begin{equation}\label{latetimes}
    \Delta V \sim \frac{R_0+R_{H}}{R_0-R_{H}} \frac{e^{2R_{H}\Delta \Phi}}{4R_{H}} ,\qquad b \rightarrow 1^-.
\end{equation}

We now relate this result to the BTZ Green function between points $(V',R',\Phi')$ and $(V,R_H,\Phi)$.  Recall that the BTZ Green function consists of a sum over image charges, such that the spikes in the Green function correspond to geodesics with initial values at the image charges.  These initial values are given by
\begin{align}\label{compare}
V_0 = V', \qquad R_0 = R', \qquad \Phi_0 = \Phi' + 2 \pi n,
\end{align}
where $n$ is any integer.  These equations also imply $\Delta V = \delta V$ and $\Delta \Phi=\Phi - \Phi' -2\pi n$.  Substituting in, we see that \eqref{latetimes} agrees exactly with the arrival times of the $n \to - \infty$ BTZ spikes \eqref{deltaVlargenegativen}, confirming that the $b \to 1^-$ geodesics are ``responsible'' for the Aretakis instability.  In Fig.~\ref{fig:plot_geodesics} we plot a selection of these trajectories.

Eqs.~\eqref{coolsolution} also display weaker, logarithmic divergences as $b\to -1^-$, corresponding to geodesics that are initially outgoing and reach a turning point at very large radius.  As opposed to the $b \to 1^-$ that spend a lot of time near the black hole, these  $b\to -1^-$ geodesics spend a lot of time near the boundary.  Expanding Eqs.~\eqref{cool1} and \eqref{cool2} (and dropping the label ``t''), we find
\begin{subequations}
\begin{eqnarray}
    \Delta V &=&  -\frac{1}{2R_{H}}\log \left|  (1+b) \frac{R_{0} -R_{H}}{4R_{H}}\right| + \mathcal{O} (1)\\
    \Delta \Phi &=& \frac{1}{R_{H}} \log \left| (1+b) \frac{R_{0} -R_{H}}{4R_{H}}\right| + \mathcal{O}(1)
\end{eqnarray}
\end{subequations}
Thus to leading order we have
\begin{equation}\label{Vweaker}
    \Delta V \sim -\frac{1}{2} \Delta \Phi_{\text{t}} ,\qquad b \rightarrow -1^{-}
\end{equation}
Making the substitutions $\Delta V = \delta V$ and $\Delta \Phi=\Phi - \Phi' -2\pi n$ (see Eq.~\eqref{compare}) now with $n>0$ to make $\Delta V$ positive, Eq.~\eqref{Vweaker} agrees precisely with the arrival times \eqref{deltaVlargen} of the weak late-time spikes $n \to \infty$ in the BTZ Green function.  Thus these weaker spikes are associated with the $b \to -1^-$ geodesics.

As discussed in detail in Sec.~\ref{sec:Ads3GF}, the Green function also contains spikes that are associated with wavefronts that bounce off the boundary.  Each such spike is in effect associated with \textrm{two} geodesics (one outgoing and one ingoing), which must be glued together by some rule determined by solving the wave equation near the boundary.  Since the Aretakis instability is visible from the above analysis of the ordinary geodesics alone, we do not study this phenomenon further.

To summarize, we have shown that geodesics originating from a fixed point $R_0>R_H$ spend arbitrarily long time outside the horizon only near the special limits $b \to 1^-$ and $b \to -1^-$.  The first limit $b \to 1^-$ corresponds to geodesics that orbit many times near the horizon and contribute wide, important spikes to the BTZ Green function at late times.  The second limit $b \to -1^-$ corresponds to geodesics that spend a long time near the timelike boundary of the spacetime and contribute narrow, unimportant spikes at late times.

\section*{Acknowledgements} We gratefully acknowledge Gary Horowitz, Don Marolf, Harvey Reall, and Bob Wald for helpful conversations. We also thank the anonymous referee for useful discussion that
improved our understanding of the origin of the on-horizon decay rate. Portions of this work were completed at the Aspen Center for Physics, which is supported by NSF grant PHY-1607611.  This work was supported in part by NSF grants PHY-1506027 and PHY-1752809 to the University of Arizona.

\appendix

\section{$\text{AdS}_3$ Green function}\label{sec:GF}
Here, we review the derivation of the Dirichlet retarded Green function in $\text{AdS}_3$ given in \cite{Danielsson:1998wt}, correcting some trivial errors and applying some identities to obtain a nice form.  The Green function equation \eqref{GF} in Poincar\'e coordinates \eqref{poincare-patch} becomes
\begin{equation}\label{boxode}
    \left(\partial_{z}^2 -\frac{1}{z} \partial_z  - \partial_{t}^2 + \partial_{x}^2  - \frac{\mu^2}{z^2}\right) G = z \delta(t-t') \delta(z-z') \delta(x - x').
\end{equation}
Mode solutions to the homogeneous equation are given by
\begin{align}\label{modes0}
    \psi_{k \omega}(z) e^{-i \omega t} e^{i k x},
\end{align}
where $\psi_{k \omega}(z)$ are related to Bessel functions \cite{Balasubramanian:1998sn}. We satisfy the Dirichlet condition by choosing
\begin{equation}\label{modes}
   \psi_{k \omega} =
   z J_{2\nu} (qz), \qquad \omega^2 = q^2 + k^2, \qquad \nu = \frac{1}{2}\sqrt{1+\mu^2}.
\end{equation}
We may use these to construct the inhomogeneous solution for $G$ as follows.  Integrating \eqref{boxode} over small interval around $t'$ replaces one of the delta functions with a jump condition,
\begin{equation}\label{eq:Gjump}
    \partial_t G |_{t \rightarrow t'^{+}} - \partial_t G_{t \rightarrow t'^{-}} = -z \delta(x-x')\delta(z-z').
\end{equation}
The retarded Green function vanishes for $t<t'$ by definition, so  \eqref{eq:Gjump} reduces to an initial condition
\begin{equation}
    \partial_t G |_{t \rightarrow t'^{+}} = -z \delta(x-x')\delta(z-z').
\end{equation}
We resolve the remaining delta functions using the relevant completeness relations
\begin{equation}
    \delta(x-x')  = \frac{1}{2\pi} \int_{-\infty}^{\infty}dk  e^{ik(x-x')}, \qquad \delta(z-z') = z' \int_{0}^{\infty} dq q J_{2\nu} (qz) J_{2\nu}(qz'),
\end{equation}
giving
\begin{equation}\label{initial}
    \partial_{t}G|_{t\rightarrow t'^{+}} = -\frac{z z'}{2\pi} \int_{-\infty}^{\infty}dk e^{ik(x-x')} \int_{0}^{\infty} dq  q J_{2\nu}(qz) J_{2\nu}(qz').
\end{equation}
The retarded Green function must vanish for $t<t'$, satisfy \eqref{initial} at $t=t'$, and be a homogeneous solution of \eqref{boxode} for $t>t'$.  In light of the homogeneous mode solutions \eqref{modes0}-\eqref{modes}, we may satisfy the latter two conditions by multiplying the right-hand side of \eqref{initial} by $-i \omega e^{-i \omega (t-t')}$.  Adjoining a step function $\Theta(t-t')$ fulfills the first condition, giving
\begin{equation}
    G=-\Theta(t-t')\frac{2zz'}{\pi} \int_{0}^{\infty} dk \int_{0}^{\infty} dq \frac{q}{\sqrt{q^2 + k^2}} \sin (\sqrt{q^2 + k^2} (t-t')) \cos (k(x-x')) J_{2\nu} (qz)J_{2\nu}(qz'),
\end{equation}
where we also use that fact that the integrand is odd in $k$ and even in $q$.  Since the Green function vanishes as we approach the boundary $z\to 0$ (or $z'\to 0$), this is the Dirichlet retarded Green function as desired.  Note that with our choice of Dirichlet boundary conditions, $G$ is conformal to the retarded propagator on the upper-half plane of Minkowski space with a mirror at infinity \cite{Dappiaggi:2016fwc} with conformal factor $zz'$.

We perform the $k$-integral using Eq.~(3.876-1) of \cite{gradshteyn2007}, giving
\begin{equation}
    G=-\Theta(t-t') \Theta(t-t' -x+x') zz'\int_{0}^{\infty} dq q J_{2\nu}(qz) J_{2\nu}( qz') J_{0} (q\sqrt{(t-t')^2 -(x-x')^2})
\end{equation}
The remaining $q$ integral is resolved with Eq.~(6.578-8) of \cite{gradshteyn2007}, giving the $\textrm{AdS}{}_3$ Green function as
\begin{align}
G_{\text{ret}}^{\text{AdS}_3} = -\Theta(t-t') \Theta(\Sigma)\begin{dcases}\frac{2}{\pi \sqrt{2 \pi} \sqrt{\sin{w}}} P_{2\nu -1/2}^{1/2} (\cos{w}) \quad &\text{if} \quad \Sigma<2 \\ \frac{-1}{\sqrt{2\pi}} \frac{\sin{(2\nu \pi)}}{\sqrt{\sinh{u}}}e^{-\frac{i\pi}{2}} Q_{2\nu -1/2}^{1/2}(\cosh{u}) \quad &\text{if} \quad \Sigma>2
\end{dcases}
\end{align}
where 
\begin{equation}
    \cosh{u} = -\cos{w}=1-\Sigma, \qquad \Sigma = \frac{(t-t')^2 - (x-x')^2 - (z-z')^2}{2zz'}.
\end{equation}
This is the form given in \cite{Danielsson:1998wt}, correcting a couple of typos.  We can simplify further using Eqs.~(14.5.17), (14.3.10), and (14.5.11) in \cite{NIST:DLMF} to obtain
\begin{align}
    G_{\rm ret}^{\text{AdS}_3} = \frac{\Theta(t-t') \Theta(\Sigma)}{\pi\sqrt{|\Sigma (\Sigma-2)|}} \begin{dcases}-\cos{(2\nu \arccos{(1-\Sigma))}} \quad &\text{if} \quad\Sigma<2 \\ \sin{(2\nu \pi)} e^{-2\nu \text{arccosh}(\Sigma-1)} \quad &\text{if} \quad \Sigma>2 \end{dcases}.
\end{align}
This form explicitly shows the inverse square root singularity at the wavefronts $\Sigma=0,2$. Another form of the result is

\begin{align}
    G_{\text{ret}}^{\text{AdS}_3} = \frac{\Theta(t-t') \Theta(\Sigma)}{2\pi\sqrt{|\Sigma (\Sigma-2)|}} \begin{dcases}- (Q(\Sigma)^{2\nu} + Q(\Sigma)^{-2\nu}) \quad &\text{if} \quad \Sigma<2 \\ 2\sin{(2\pi \nu)} Q(\Sigma)^{-2\nu} \quad &\text{if} \quad \Sigma>2 \end{dcases},
\end{align}
where $Q(\Sigma) = |1-\Sigma| + \sqrt{\Sigma (\Sigma-2)}$.  The large-$\Sigma$ behavior is
\begin{equation}\label{Gadslate}
    G_{\text{ret}}^{\text{AdS}_3} \sim \sin (2\pi \nu) \frac{4^{-\nu}}{\pi} \Sigma^{-2\nu-1},\qquad \Sigma \rightarrow \infty.
\end{equation}

\section{Massless axisymmetric perturbations: Aretakis' method}\label{sec:Aretakis}

 Using the technique  employed originally  by Aretakis for extreme black holes in four dimensions \cite{Aretakis:2010gd,Aretakis:2011hc}, in this appendix we show that massless linear perturbations arising from axisymmetric data with support extending to the horizon conserve a charge on the extremal BTZ horizon.  As in the four dimensional case, the associated conservation law prevents derivative decay, and higher-order derivatives grow polynomially. Assuming decay of the perturbation itself, we obtain rates for the derivatives by a hierarchy argument. 

To begin we express the massless wave operator in ``radially inverted'' ingoing cororating coordinates $(V,\rho,\Phi)$, where $\rho=R_H/R$,
\begin{equation}
\label{eq:wave coro R-inverted}
\Box\psi = (\rho^2-1)f[\psi] + \pd_V \left(1-2\rho \pd_\rho\right)\psi - \pd_\Phi \left((1+\rho^2)- \frac{\rho}{R_H}\pd_\Phi\right) \psi.
\end{equation}
Here, $f$ is a 2nd-order differential operator having smooth  coefficients in a neighborhood of $\rho=1$ (the horizon) whose precise form is irrelevant for the arguments to follow. 
Integrating the wave equation on a cylindrical section of the horizon $\rho=1$ extending from an initial time $V_1$ to some final time $V_2$ leads to the conservation law 
\begin{equation}\label{eq:H global}
    \int_{S^1} \int_{V_1}^{V_2} dV  \pd_V (1-2\rho\pd_\rho)\psi \vert_{\rho=1} \,  =0,
\end{equation}
where  we have used that $\int_{S^1} \pd_\Phi \psi = 0$ for sufficiently smooth $\psi$.
The conservation law is stated locally as
\begin{equation}\label{eq:H cons}
 \pd_V H \vert_{\mathcal H} = 0,
\end{equation}
where
\begin{equation}\label{eq:H def}
H \equiv \int_{S^1} (1-2\rho\pd_\rho) \psi \vert_{\rho=1}.
\end{equation}
  The conservation law implies that if $\psi$ decays to zero on the horizon as $V\to\infty$, the radial derivative must asymptote to a constant.  Compactly,
\begin{equation}\label{eq:Darth v}
 [\psi_{\mathcal H} \to 0  \quad \text{and} \quad  \eqref{eq:H cons}] \implies \pd_\rho \psi_{\mathcal H} \to -\frac{H}{2}.
\end{equation}
Furthermore, if both $\psi$ and its tangential $V$ derivatives along the horizon decay asymptotically as $V\to\infty$, then higher-order radial derivatives grow at rates 
\begin{equation}\label{eq:div blowup}
\pd_\rho^n \psi_{\mathcal H} \sim c_n(H)\, V^{n-1}, \qquad V\to \infty,
\end{equation}
where $c_n$ is an order one constant depending on the specific choice initial 
data through $H$.

To demonstrate the blow-up of the first derivative,  take the radial derivative of the wave equation $\pd_\rho \Box \psi=0$, pull back to the horizon, and average over the circular cross section:
\begin{equation}\label{eq:berg}
0
= \int_{S^1} \Big[8R_H \pd_\rho\psi+\pd_V \left(\psi-4 \pd_\rho\psi-2\pd_\rho^2\psi\right)\Big]_{\mathcal H}.
\end{equation}
Using our  decay assumptions and results above for the horizon rates $\psi\to 0$, $\pd_V \psi \to 0$, and $\pd_\rho \psi \to -H/2$ as $V\to \infty$,  integration  of \eqref{eq:berg} gives  linear growth
\begin{equation}\label{eq:D2 psi}
 \pd_\rho^2 \psi_{\mathcal H} \sim -4R_H\, H V , \qquad V \to \infty.
\end{equation}
The derivation of the rates for higher derivatives proceeds analogously, and straightforward induction yields \eqref{eq:div blowup}.

\section{Mode approach}\label{sec:modes}

\subsection{Preliminaries}
 
The starting point for our mode calculations is the introduction of corotating, constant proper volume coordinates given by
\begin{equation}\label{eq:vc coords}
    T=T,\qquad y=\frac{2R_{H}^2}{R^2 - R_{H}^2},\qquad  \eta = X-  T.
\end{equation}
In these coordinates the separated mode solutions resemble those found for perturbations of $\mathrm{AdS}_2$ with an electic field at small frequencies, whose properties have been exhaustively analysed in previous studies of the Aretakis instability \cite{Gralla:2018xzo}. 
Decomposing into modes, the field takes the form of a Fourier-Laplace integral
\begin{equation}\label{eq:psi modes}
    \psi =  \frac{1}{2\pi}\sum_{m=-\infty}^{\infty}e^{i m \eta} \int_{-\infty}^{\infty}  d\omega\, e^{-i\omega T}  \psi_{m \omega} (y) .
\end{equation}
For Dirichlet data, the field modes $\psi_{m \omega}$ are determined by a radial convolution integral containing mode decomposed initial data and an integral kernel called the ``transfer function''. The range of the integral spans the support of the data, which we choose to be bounded away from the event horizon and infinity. Since we are considering generic mode-evolution of this data, we choose to work directly with the transfer function, denoted here as $g_{m}(y,y';\omega)$. It satisfies the inhomogeneous equation
\begin{equation}\label{eq:gmw}
g_{m}''(y,y';\omega) + \left(\frac{\omega^2}{16 R_{H}^2} + \frac{\omega (\omega+2m)}{8 R_{H}^2 y} + \frac{h(1-h)}{y^2} \right)g_{m}(y,y';\omega) = \frac{1}{4R_{H}^2} \delta(y-y'),
\end{equation}
where prime denotes ordinary $y$-differentiation. Here again, $h$ is as defined in \eqref{eq:h}. 

The proceeding arguments may be understood directly from the BTZ wave equation and Kirchhoff representation of the mode decomposed retarded propagator 
\begin{equation}\label{eq:Gf modes}
    G=\frac{1}{2\pi} \sum_{m=-\infty}^\infty e^{im(\eta-\eta')} \int_{-\infty}^\infty d\omega\, g_{m}(y,y';\omega) e^{-i\omega(T-T')} .
\end{equation}

To ensure causal evolution, we choose the transfer function corresponding to the retarded solution. This choice dictates our selection of homogeneous radial functions subject to boundary conditions corresponding to ingoing waves at the horizon $R_{\text{in}}$ and Dirichlet falloff infinity $R_{\text{D}}\sim y^h$ as  $y\to 0$. These solutions exist provided $\mu^2>-1$ \cite{Ishibashi:2004wx}. Imposing a $C^0$ match of the two homogeneous solutions at the support of the delta function allows the transfer function to be written as 
\begin{equation}\label{eq:transfer}
g_{m}(y,y';\omega) = \frac{1}{4R_{H}^2}\frac{R_{\text{in}}(y_{>}) R_{\text{D}}(y_{<})}{\boldsymbol{W}[R_{\text{in}}(y'), R_{\text{D}}(y')]},
\end{equation}
where $y_{<} = \text{min}(y,y')$ and $y_{>} = \text{max}(y,y')$. Here we have introduced the conserved Wronskian of the two solutions $\boldsymbol{W} \left[ R_{\text{in}}, R_{\text{D}}\right]=R_{\text{in}} R_{\text{D}}' - R_{\text{D}} R_{\text{in}}'$.
We express the homogeneous solutions in  terms of Whittaker functions 
as
\begin{equation}\label{eq:in and up}
R_{\text{in}}= W_{iq, \nu} \left(\frac{-i\omega y}{2R_{H}}\right), \qquad R_{\text{D}}= M_{-iq, \nu} \left(\frac{i\omega y}{2R_{H}}\right),
\end{equation}
where we have defined
\begin{equation}\label{eq:q def}
q = \frac{1}{2R_{H}}\left(m + \frac{ \omega}{2}\right).
\end{equation}
These solutions are linearly independent provided $\nu$ is non-integer. This special case will not be considered here.
Using Eq.~(13.14.27) of Ref.~\cite{NIST:DLMF} for the Wronskian of these two functions, we find that \eqref{eq:transfer} can be written as
\begin{equation}\label{eq:tf}
    g_{m}(y,y';\omega) = \frac{(-1)^h}{2R_{H}}  \frac{\Gamma(h-iq)}{i \omega \Gamma(2h)} 
    \, W_{iq, \nu} \left( \frac{-i\omega y_{>}}{2R_{H}}\right) M_{-iq, \nu} \left(\frac{i\omega y_{<}}{2R_{H}} \right).
\end{equation}

The coordinates $(T,y,\eta)$ do not extend to the horizon of the black hole, and thus fail to characterize Aretakis' instability. For a suitable extension to $R_H$ we now readopt the ``horizon  coordinates'' $V$ and $\Phi$ (see Eq.~\eqref{horizon-coordinates}) and invert $y$, 
\begin{equation}\label{eq:btz barred}
      V=T+{R}_{*}, \qquad \xh = \frac{1}{y},\qquad \Phi = \eta+{\Phi}_{\sharp}.
\end{equation}
Explicitly, ${R}_{*}$ and ${\Phi_{\sharp}}$ are given in terms of $\xh$ by
\begin{subequations}
\begin{align}\label{eq:Rs Phis}
    {R}_{*} &=  -\frac{1}{4R_{H}}\left( \frac{\sqrt{1+2\xh}}{\xh} + \log \left(\frac{1+\xh+\sqrt{1+2\xh}}{\xh} \right) \right),\\ 
    {\Phi}_{\sharp} &=  \frac{1}{2R_{H}}\log \left( \frac{1+\xh+\sqrt{1+2\xh}}{\xh}\right).
\end{align}
\end{subequations}
In these coordinates the horizon sits at $\xh=0$. On the horizon, $\pd_V$ is tangent and $\Phi$ labels generators. While at infinity, $ V=T$ and $\Phi=\eta$.
From the $T-T'$ and $\eta-\eta'$ dependence in the phases $e^{-i\omega(T-T')}$ and $e^{im(\eta-\eta')}$ of the mode decomposition \eqref{eq:Gf modes}, we see that, under the coordinate transormation \eqref{eq:btz barred}, the radial Green function $g_{m\omega}$ is transformed by
\begin{equation}\label{eq:gtog}
    g_{m} \to e^{-im({\Phi}_{\sharp} - {\Phi}'_{\sharp}) + i \omega ({R}_{*} - {R}_{*}')} g_{m}.
\end{equation}
Both the primed and unprimed phases may be determined from the expression
\begin{equation}\label{eq:exp Rs simp}
    e^{i\omega  R_{*} - i m  \Phi_{\sharp}} =\left( \frac{\xh}{1+\xh+\sqrt{1+2\xh}} \right)^{iq}\, \exp{ \Bigg(-\frac{i\omega}{4R_{H}}\frac{\sqrt{1+2\xh}}{\xh} \Bigg)}
\end{equation}
by complex conjugating at the primed point.
    

\subsection{Modes at late times near the horizon}
To obtain the late-time behavior of a fixed $m$ mode, we apply the asymptotic theory of Fourier-Laplace transforms \cite{Doetsch1974} wherein the functional behavior as $T\to\infty$ is determined by the leading-order non-analytic term in an asymptotic expansion of the Laplace transform about its uppermost singular point in the complex $\omega$ plane. 

Inspection of \eqref{eq:tf} reveals that the transfer function has its uppermost singular point at $\omega=0$. To determine the late-behavior near the horizon, we expand $g_{m\omega}$ in an asymptotic series about $\omega=0$, keeping $\omega/\xh$ held fixed. This limit motivates the definition of the  near-far late-time transfer function
\begin{equation}\label{eq:gnf}
    g^{\rm nf,lt}_m := g_m(\omega\to0)\qquad \text{fixing} \quad \omega/\xh,
\end{equation}
which excludes all terms analytic in $\omega$ with the exception of those involving $\omega/\xh$.
In the late-time limit the $M$ function simplifies via Eq.~(13.14.14) of Ref.~\cite{NIST:DLMF}, and we find
\begin{equation}\label{eq:gw}
    g_{m}^{\rm nf,lt} = 
   -\frac{\Gamma(h-iq_0)}{4R_H^2\Gamma(2h)}\,
   \Big[ \xh^{iq_0}\xh{'}^{-h}\, 
e^{2iR_Hq_0\Phi_\sharp'}\Big]
    \left(-i\omega/2R_H\right)^{h-1} W_{iq_0,\nu}\left(-\frac{i\omega}{2R_H \xh}\right) e^{ -\frac{i\omega}{4 R_H \xh}}.
\end{equation}
Here we have 
defined 
\begin{equation}\label{eq:q0}
    q_0 = q\vert_{\omega=0}=\frac{m}{2R_H}.
\end{equation}
To perform the inverse transform, we use the identity Eq.~(9) in Sec.~5.20, of Ref \cite{bateman1954tables}.
With this, the late-time result, stated terms of the shifted time coordinate
\begin{equation}\label{eq:dV}
    \delta V=(V-V')
\end{equation}
introduced previously, is
\begin{equation}\label{eq:glt}
    g_m^{\rm nf,lt}(\xh,\xh';\delta V)
    =- \frac{\xh{'}^{-h} e^{2iR_H q_0 \Phi_\sharp'}\Gamma(h-iq_0)}{4R_H^2\Gamma(2h)\Gamma(1-h-iq_0)} \Big[
    (2R_H\delta V)^{-h-iq_0}(1+2R_H\,\xh\delta V)^{-h+iq_0}
    \Big]\, .
\end{equation}

We conclude this section with a few remarks regarding our result for $g_m^{\rm nf,lt}$.  On the horizon, the decay of a fixed $m$-mode is given by $\delta V^{-h}$.  Higher transverse derivatives  $\pd_R^n\vert_{\mathcal H}$ exhibit the Aretakis instability, growing at rate $\delta V^{-h+n}$. These decay and instability rates are consistent with those determined from the sum over images using the method of images above in the main text. However, at large $m$, the magnitude of $g_m^{\rm nf,lt}$ grows as $m^{2h-1}$, indicating that the mode sum does not converge pointwise (recall that $h\geq 1/2$).  The high-$m$ regime of the transfer function probes the roughest part of the initial data's $\Phi$-dependence.  If the data is sufficiently regular, with differentiability exceeding $2h-1$, then its Fourier coefficients will decay in $m$ at a faster rate than the $m$-growth exhibited by the coefficients in the transfer function, such that formal application of the Kirchhoff integral \eqref{kirchhoff} provides a finite result for the field.  However, the field will then be less regular than the initial data, in contradiction with rigorous results of Warnick \cite{Warnick:2012fi}.  Thus our results for each $m$-mode cannot be straightforwardly promoted to results about the whole field, indicating some nonuniformity in the late-time and large-$m$ limits.  Nevertheless, the rates predicted by the mode expansion do agree with those of the image sum.  We hope to explore these issues further in future work.

\subsection{Non-periodic limit - $\text{AdS}_3$}

The late-time behavior \eqref{eq:glt} of each angular mode displays growth of transverse derivatives, consistent with the full extremal BTZ spacetime possessing the Aretakis instability.  However, the angular modes also apply to the AdS${}_3$ spacetime, which has no such instability.  From the perspective of the mode approach, the only difference between AdS${}_3$ and BTZ is that the latter admits only a subset of modes compatible with the periodicity of $\Phi$ (i.e., $m$ is continuous for AdS${}_3$ and quantized for BTZ).  That is, formally speaking we have
\begin{align}
    G^{\text{nf,lt}}_{\text{BTZ}} & = \sum_{m=-\infty}^{\infty} e^{im(\Phi - \Phi')} g_{m}^{\text{nf,lt}}, \label{GnfltBTZ}\\
    G^{\text{nf,lt}}_{\text{AdS}_3} & = \frac{1}{2\pi} \int^{\infty}_{-\infty} e^{im(\Phi - \Phi')} g_{m}^{\text{nf,lt}} dm. \label{GnfltAdS3}
\end{align}
We have already noted that the mode sum \eqref{GnfltBTZ} does not converge due to $g_m^{\rm nf,lt}$ behaving as $m^{2h-1}$ at large $m$, and the integral \eqref{GnfltAdS3} is likewise divergent.  However, the following formal manipulation assigns it a value that agrees with $\mathrm{ AdS{}}_3$ expectations.  Using Eq.~\eqref{eq:glt} and expressing the integral \eqref{GnfltAdS3} in terms of $q_0$ defined in Eq.~\eqref{eq:q0}, we have 
\begin{align}\label{pre-magic}
    G_{\text{AdS}_3}^{\text{nf,lt}} = -\frac{ \left[Y'(2R_{H} \delta V)(1+2R_{H}Y \delta V ) \right]^{-h}}{2R_{H}\Gamma(2h)}  \int_{-\infty}^{\infty} dq_{0}\, \frac{\Gamma(h-iq_{0})}{\Gamma(1-h-iq_{0})} e^{i q_{0} a},
\end{align}
where $a = \left[2R_{H}(\Phi - \Phi' + \Phi_{\sharp}') + \log \left( \frac{1+ 2R_{H} Y\delta V }{2R_{H} \delta V}\right)\right]$.

Assuming the generic case $2h \notin \mathbb{Z}^+$, the integrand contains poles at $q_{0}=-i(h+n)$ with $n \in \mathbb{Z}^+$.  By Cauchy's theorem we may express the integral as a sum over these poles, together with a contribution from a semicircular arc at large radius in the lower-half plane.  This arc contribution does \textit{not} vanish, and in fact it is infinite on account of the behavior $q_0^{2h-1}$ of the integrand at large $|q_0|$.  Ignoring this infinite contribution provides a natural regulator that leaves a finite result for the integral, namely the residue sum.  This sum can be done in closed form using the formula
\begin{equation}
    \sum_{n=0}^{\infty} \frac{(-1)^{n}e^{(n+h)a}}{\Gamma(n+1) \Gamma(1-2h-n)}  = \frac{e^{ah}(1-e^{a})^{-2h}}{\Gamma(1-2h)}.
\end{equation}
The result for $G_{\text{AdS}_3}^{\text{nf,lt}}$, regulated by dropping the arc contribution, is then
\begin{align}\label{magic}
    G_{\text{AdS}_3}^{\text{nf,lt}} & = -\frac{Y'^{-h} e^{2R_{H}h(\Phi - \Phi' + \Phi_{\sharp}') }}{2R_{H} \Gamma(1-2h) \Gamma(2h)}\left[ 1-e^{2R_{H}(\Phi - \Phi' + \Phi_{\sharp}')} \left(
    Y + \frac{1}{2R_H\delta V}\right)\right]^{-2h} \left(2R_{H} \delta V \right)^{-2h}.
\end{align}
The dependence on $Y \delta V$ in the mode result \eqref{eq:glt} has now disappeared, and correspondingly Eq.~\eqref{magic} shows no Aretakis instability (all transverse derivatives decay).  In fact, the second term in square brackets is subleading in the limit we consider ($\delta V \to \infty$ fixing $Y \delta V$),\footnote{The corresponding term appeared in a phase in Eq.~\eqref{pre-magic} and could not be dropped at that stage} so that to leading order we have simply 
\begin{align}
    G_{\text{AdS}_3}^{\text{nf,lt}} & = -\frac{Y'^{-h} e^{2R_{H}h(\Phi - \Phi' + \Phi_{\sharp}') }}{2R_{H} \Gamma(1-2h) \Gamma(2h)}\left(2R_{H} \delta V \right)^{-2h}.
\end{align}

Although there are many unresolved issues with convergence and regulators, the discussion in this appendix supports the following general picture: Each angular mode in BTZ or $\mathrm{AdS}_3$ displays the Aretakis instability, but these are only promoted to an instability arising from compactly supported initial data in the BTZ spacetime, where a subset of modes is selected out by the periodic identification.  This is consistent with the simple idea that a single angular mode in BTZ has compact spatial support, while a corresponding mode in the AdS${}_3$ spacetime does not.

\bibliography{MyReferences.bib}
\end{document}